\def\lsim{\mathrel{\rlap{\lower4pt\hbox{\hskip1pt$\sim$}}
    \raise1pt\hbox{$<$}}}                
\def\gsim{\mathrel{\rlap{\lower4pt\hbox{\hskip1pt$\sim$}}
    \raise1pt\hbox{$>$}}}                
\def\AA{\buildrel _{\circ} \over {\mathrm{A}}}

\documentclass{emulateapj}
\usepackage{psfig}
\usepackage{apjfonts}

\slugcomment{Accepted for publication in ApJ}
\shorttitle{Color-Magnitude Relations of Active and Non-Active Galaxies}
\shortauthors{Xue et al.}

\begin{document}  

\title{Color-Magnitude Relations of Active and Non-Active Galaxies in the {\it Chandra} Deep Fields: High-Redshift Constraints and Stellar-Mass Selection Effects}
\author{
Y. Q. Xue\altaffilmark{1,2},  
W. N. Brandt\altaffilmark{1,2},  
B. Luo\altaffilmark{1,2}, 
D. A. Rafferty\altaffilmark{1,3},   
D. M. Alexander\altaffilmark{4},  
F. E. Bauer\altaffilmark{5,6},  
B. D. Lehmer\altaffilmark{4,7,8}, 
D. P. Schneider\altaffilmark{1}, and  
J. D. Silverman\altaffilmark{9,10}
}
\altaffiltext{1}{Department of Astronomy and Astrophysics, Pennsylvania State University, University Park, PA 16802, USA; xuey@astro.psu.edu}
\altaffiltext{2}{Institute for Gravitation and the Cosmos, Department of Physics, Pennsylvania State University, University Park, PA 16802, USA}
\altaffiltext{3}{Current address: Leiden Observatory, Leiden University, Oort Gebouw, P.O.Box 9513 RA, Leiden, The Netherlands}
\altaffiltext{4}{Department of Physics, Durham University, Durham, DH1 3LE, UK}
\altaffiltext{5}{Space Science Institute, 4750 Walnut Street, Suite 205, Boulder, Colorado 80301}
\altaffiltext{6}{Pontificia Universidad Cat\'{o}lica de Chile, Departamento de Astronom\'{\i}a y Astrof\'{\i}sica, Casilla 306, Santiago 22, Chile}
\altaffiltext{7}{The Johns Hopkins University, Homewood Campus, Baltimore, MD 21218, USA}
\altaffiltext{8}{NASA Goddard Space Flight Centre, Code 662, Greenbelt, MD 20771, USA}
\altaffiltext{9}{ETH Zurich, Institute of Astronomy, Department of Physics, Wolfgang-Pauli-Strasse 16, 8093 Zurich, Switzerland}
\altaffiltext{10}{Institute for the Physics and Mathematics of the Universe (IPMU), University of Tokyo, Kashiwanoha 5-1-5, Kashiwa-shi, Chiba 277-8568, Japan}

\begin{abstract}
We extend color-magnitude relations for moderate-luminosity X-ray AGN hosts and non-AGN galaxies through the galaxy formation epoch (\mbox{$z\approx 1$--4}) in the {\it Chandra} Deep Field-North and {\it Chandra} Deep Field-South (\mbox{CDF-N} and \mbox{CDF-S}, respectively; jointly CDFs) surveys.
This study was enabled by 
the deepest available X-ray data from the \mbox{2~Ms} CDF surveys as well as complementary ultradeep multiwavelength data in these regions.
%
%
We utilized analyses of color-magnitude diagrams (CMDs) to assess the role of moderate-luminosity AGNs in galaxy evolution.
First, we confirm some previous results and extend them to higher redshifts, finding, for example, that
(1) there is no apparent color bimodality (i.e., the lack of an obvious red sequence and blue cloud) for AGN hosts from \mbox{$z\approx 0-2$}, 
but non-AGN galaxy color bimodality exists up to \mbox{$z\approx 3$} and the relative fraction of red-sequence galaxies generally increases as the redshift decreases (consistent with a blue-to-red migration of galaxies);
(2) most AGNs reside in massive hosts and the AGN fraction rises strongly toward higher stellar mass, up to \mbox{$z\approx 2$--3}; and
(3) the colors of both AGN hosts and non-AGN galaxies become redder as the stellar mass increases, up to \mbox{$z\approx 2$--3}.
Second, we point out that, in order to obtain a complete and reliable picture,
it is critical to use mass-matched samples to examine color-magnitude relations of AGN hosts and non-AGN galaxies.
We show that for mass-matched samples up to \mbox{$z\approx 2$--3},  
AGN hosts lie in the same region of the CMD as non-AGN galaxies; 
i.e., there is no specific clustering of AGN hosts in the CMD around the red sequence, the top of the blue cloud, or the green valley in between.
The AGN fraction (\mbox{$\approx 10$\%}) is mostly independent of host-galaxy color, providing an indication of the duty cycle of supermassive black hole growth in typical massive galaxies.
These results are in contrast to those obtained with non-mass-matched samples
where there is apparent AGN clustering in the CMD and
the AGN fraction generally increases as the color becomes redder.
We also find, for mass-matched samples, that the star-formation rates of AGN hosts are typically
a factor of \mbox{$\approx 2$--3} larger than those of non-AGN galaxies at \mbox{$z\approx 0$--1},
whereas this difference diminishes at \mbox{$z\approx 1$--3}.
With mass-selection effects taken into account, we find that almost all the results obtained in this work
can be reasonably explained by two main ingredients, color-mass correlation (i.e., 
X-ray AGNs preferentially reside in massive galaxies that generally tend to be redder than less-massive galaxies) 
and passive or secular evolution of galaxies.
%
Our results show that the presence of moderate-luminosity AGN activity does not have a significant effect on the colors of galaxies and 
thus tightly constrain any effects from moderate-luminosity AGN feedback upon color-magnitude properties over the \mbox{$\approx 80$\%} of cosmic time during which most of galaxy formation occurred. 
\end{abstract}


\section{Introduction}\label{intro}

Deep multiwavelength surveys have greatly refined our understanding of the cosmic star-formation and mass-assembly history.
Measurements of galaxy luminosity functions at multiple wavelengths have provided a reasonably tight constraint
on the evolution of the space density of the galaxy star-formation rate (SFR): the cosmic star formation history
is known to within \mbox{$\approx 30$--50\%} up to redshifts of \mbox{$z\approx 1$} and within a factor of \mbox{$\approx 3$} at \mbox{$z\approx 1$--6} (e.g., Hopkins 2004).
It has been established that the galaxy SFR density has a broad maximum at \mbox{$z\approx 1$--4},
followed by a sharp drop from \mbox{$z\approx 1$} to \mbox{$z\approx 0$} (e.g., Dickinson et al. 2003a;
Giavalisco et al. 2004) and a fast decrease beyond \mbox{$z\approx 3$--4} (see, e.g., Hopkins \& Beacom 2006; Wilkins et al. 2008).
It now seems clear that the bulk of present stellar mass was formed over a critical epoch at \mbox{$z\approx 1$--4} (e.g., Hopkins \& Beacom 2006).

Star-formation and active galactic nucleus (i.e., AGN) activity in galaxies are often found to go hand-in-hand (e.g., Page et al. 2001; Alexander et al. 2005b; Netzer et al. 2007; Silverman et al. 2009).
Both AGN activity and star formation peak in about the same redshift range (\mbox{$z\approx 1.5-2$}), 
and they undergo a similar decline below \mbox{$z\approx 1$} (e.g., Cowie et al. 2003; Merloni et al. 2004; Silverman et al. 2008a).
Mounting evidence has shown that supermassive black holes (SMBHs) centered in galaxies must play an important role in galaxy evolution.
The close connection between AGNs and their hosts is most strikingly shown by the tight correlations between
the masses of SMBHs and the properties of their host-galaxy bulges (e.g., H\"{a}ring \& Rix 2004; Ferrarese \& Ford 2005; G\"{u}ltekin et al. 2009).
Recent theoretical models propose that AGN feedback is an important ingredient in understanding the interconnection between AGNs
and their hosts (e.g., Silk \& Rees 1998; Di Matteo et al. 2005; Bower et al. 2006; Croton et al. 2006; Fabian, Vasudevan, \& Gandhi 2008).
Two main AGN feedback modes are thought to be significant.
In the ``radiative mode'',
the initially shrouded luminous AGN expels the obscuring gas (that feeds both the SMBH and star formation)
and thus quenches star formation (e.g., Hopkins et al. 2005; Springel et al. 2005).
The ``kinetic mode'' invokes AGN heating to prevent hot gas from cooling and falling into a galaxy to form stars (e.g., Croton et al. 2006), which is generally thought to be important in the more massive halos and occurs at much smaller accretion rates than that of the radiative-mode feedback.
To investigate the close connections between star formation and AGN activity and 
thus fully understand galaxy formation and evolution, 
a sensible approach is to study both AGNs and their hosts.

Galaxy colors are directly related to the star-formation, dust, and metal-enrichment history of galaxies and thus provide important constraints for models of galaxy formation and evolution.
The color-magnitude diagram (CMD; plot of rest-frame $U-V$/$U-B$ colors vs. $V$-band/$B$-band absolute magnitudes) has proven to be one effective tool for exploring the role of AGNs in galaxy evolution.
Previous CMD studies have obtained a number of results, e.g.:
(1) The galaxy color bimodality (i.e., separation of galaxies into the red sequence and the blue cloud) has been clearly seen in the CMD, 
while no color bimodality of \mbox{X-ray} AGN hosts appears to exist (e.g., B\"{o}hm \& Wisotzki 2007; Nandra et al. 2007; Silverman et al. 2008b).    
(2) The clustering of AGNs in the CMD is distinct. It has been shown that 
\mbox{X-ray}/optically-selected AGNs reside in the red sequence, the top of the blue cloud, and the green valley in between (e.g., Martin et al. 2007; Nandra et al. 2007; Rovilos \& Georgantopoulos 2007; Westoby et al. 2007);
it has also been found that radio AGNs
preferentially lie on the red sequence, \mbox{X-ray} AGNs lie in the green valley, and
IR-selected AGNs are found in somewhat bluer hosts than \mbox{X-ray} AGNs (Hickox et al. 2009).
(3) The broad distribution of host-galaxy colors of \mbox{X-ray} selected moderate-luminosity AGNs is dependent on the strong color evolution of luminous
(\mbox{$M_{\rm V}<-20.7$}) galaxies and the influence of enhanced AGN activity in \mbox{$\approx 10$} Mpc large-scale structures (Silverman et al. 2008b).
(4) Morphologies of \mbox{X-ray} selected moderate-luminosity AGNs
reveal no close connection between major mergers and AGN activity, 
and are consistent with minor interactions and/or secular evolution (e.g., Grogin et al. 2005; Georgakakis et al. 2008; Silverman et al. 2008b). 
(5) \mbox{X-ray} AGNs preferentially reside in luminous bulges (e.g., Nandra et al. 2007; Silverman et al. 2008b).
However, we note that these CMD results are limited to redshifts below \mbox{$\approx 1.4$} and relatively bright sources (usually \mbox{$m_{\rm R}\leq 24$}), primarily due to data-depth limitations.
A natural follow-up question is what further insights can be gained by extending CMD studies through the galaxy formation epoch when star-formation and AGN activity peaked?

As data from exceptionally deep multiwavelength surveys have become available in the \mbox{2~Ms} {\it Chandra} Deep Fields (see Brandt \& Hasinger 2005 and Brandt \& Alexander 2010 for reviews of deep extragalactic \mbox{X-ray} surveys),
it is now feasible to push CMD studies to higher redshifts. 
In this paper, we aim to assess the role of AGNs in galaxy evolution by means of CMD analyses through the galaxy formation epoch, using the \mbox{2~Ms} CDF surveys (e.g., Alexander et al. 2003; Luo et al. 2008) as well as superb 
complementary ultradeep multiwavelength data. 
A key aspect of this paper is the use of reliable stellar masses and the assessment of possible stellar-mass selection effects,
because stellar mass is likely the most fundamental observable parameter for understanding the properties of galaxies and there are suggestions of stellar-mass biases in CMD works (e.g., Silverman et al. 2009).
This paper is structured as follows: 
\S~\ref{data} describes the multiwavelength data used in this work;
\S~\ref{properties} shows the derivations of source physical properties as well as AGN identification;
\S~\ref{sec:sample} details sample construction;
\S~\ref{results} gives the results obtained by this work, 
where we critically examine
issues such as color bimodality, AGN clustering in the CMD, and evolutionary trends in the colors of AGN hosts and non-AGN galaxies, with mass-selection effects taken into account;
and finally, \S~\ref{summary} presents the conclusions and summary. 
Throughout this paper, all absolute magnitudes quoted are based upon the Vega magnitude system;
\mbox{X-ray} luminosities are absorption-corrected (i.e., intrinsic) and quoted in the \mbox{0.5--8}~keV full band; 
and a cosmology of $H_0=70.5$ km s$^{-1}$ Mpc$^{-1}$, $\Omega_{\rm M}=0.274$, and $\Omega_\Lambda=0.726$ derived from the five-year {\it WMAP} observations (Komatsu et al. 2009) is adopted.

\section{Multiwavelength Data}\label{data}

\subsection{Source Catalogs}\label{cats}

{\it \mbox{X-ray} catalogs:}
We made use of the \mbox{2~Ms} CDF point-source catalogs.
The \mbox{CDF-N} catalog (Alexander et al. 2003) consists of 582 \mbox{X-ray} point sources 
(503 in the main {\it Chandra} source catalog
and 79 in the supplementary optically bright {\it Chandra} source catalog).
The \mbox{CDF-S} catalog (Luo et al. 2008) consists of 578 \mbox{X-ray} point sources 
(462 in the main {\it Chandra} source catalog,
86 in the supplementary \mbox{CDF-S} plus \mbox{E-CDF-S} {\it Chandra} source catalog,
and 30 in the supplementary optically bright {\it Chandra} source catalog).

{\it Optical/UV/IR catalogs:}
For the \mbox{CDF-N}, we used the Hawaii \mbox{HDF-N} optical and NIR catalog ($U,B,V,R,I,z^\prime,HK^\prime$; Capak et al. 2004) 
as the base catalog that has a total of 48,858 sources.
This base catalog is sufficiently deep and complete to $m_{\rm R}\leq 26$.
We cross-matched four other available optical/UV/IR catalogs to the Hawaii \mbox{HDF-N} catalog and kept only sources that have counterparts in the Hawaii \mbox{HDF-N} catalog. 
Those catalogs are
(1) the GOODS-N ACS and IRAC catalogs (ACS F435W, F606W, F775W, F850LP, and IRAC 3.6 $\mu$m, 4.5 $\mu$m, 5.8 $\mu$m, 8.0 $\mu$m; Dickinson et al. 2003b);\footnote{See http://archive.stsci.edu/pub/hlsp/goods/catalog\_r2/.}$^,$\footnote{See http://ssc.spitzer.caltech.edu/legacy/goodshistory.html.}
(2) the GALEX \mbox{HDF-N} deep imaging survey catalog (NUV, FUV; GALEX Release 4 Data\footnote{See http://galex.stsci.edu/GR4/.\label{ft-galex}}); (3) the ACS GOODS-N region $K_{\rm s}$ ($<24.5$) catalog (Barger, Cowie, \& Wang 2008); and (4) the \mbox{GOODS-N} MIPS 24 $\mu$m catalog (Dickinson et al. 2003b). 
In this work we adopted the aperture-corrected photometry as detailed in the original references.

For the \mbox{CDF-S}, we combined three catalogs to produce a base catalog: (1) the MUSYC $BVR$-detected optical catalog ($U,B,V,R,I,o3,z$; Gawiser et al. 2006); (2) the COMBO-17 optical catalog ($U,B,V,R,I$ + 12 medium-band filters; Wolf et al. 2004, 2008); and (3) the GOODS-S MUSIC catalog ($U$, F435W, F606W, F775W, F850LP, $J,H,K_{\rm s}$, and IRAC; Grazian et al. 2006). All MUSYC sources and all the unique COMBO-17 and MUSIC sources (i.e., COMBO-17 and MUSIC sources that did not match to a MUSYC source, with a matching radius of 0.5\arcsec) were kept. With this approach, the base catalog has a total of 100,318 sources.
This base catalog is also sufficiently deep and complete to $m_{\rm R}\leq 26$.
We then cross-matched four other available catalogs to the base catalog and kept only sources that have counterparts in the base catalog. Those catalogs are (1) the MUSYC NIR catalog ($J,H,K_{\rm s}$; Taylor et al. 2009b); (2) the SIMPLE IRAC catalog (Damen et al. 2010); (3) the GALEX \mbox{CDF-S} deep imaging survey catalog (NUV, FUV; GALEX Release 4 Data); and (4) the \mbox{GOODS-S} MIPS \mbox{24 $\mu$m} catalog (Dickinson et al. 2003b).
Again, we adopted the aperture-corrected photometry for analysis.

\subsection{Cross-Matching}\label{matching}
Two cross-matching methods were used: closest-counterpart matching and likelihood-ratio matching. Closest-counterpart matching, as the name suggests, simply assigns the closest angular match as the counterpart (given some maximum matching radius); the separation between two sources is the sole criterion for this method (so it is easily implemented and executes quickly). Closest-counterpart matching performs acceptably (i.e., with a low false-match probability\footnote{We estimated the false-match probability by shifting the coordinates of base sources in RA and DEC by $\pm 5$\arcsec \hspace{0.35mm} 
and re-correlating with sources from other catalogs.}), provided that an appropriate maximum-matching radius is adopted (see, e.g., Table~\ref{match}). 
Here, closest-counterpart matching was used for the cross-matching between the base catalogs and other optical/UV/IR catalogs.

\begin{table}[ht]
\caption{Cross-Matching Results for the Central $r_{\rm encircled}=$8\arcmin \hspace{0.5mm} Areas}
\begin{tabular}{lcccc}\hline\hline
        & Matching & Median & False & Matching \\
Catalog & Radius (\arcsec) & Offset (\arcsec) & Rate & Method \\
(1) & (2) & (3) & (4) & (5) \\ \hline
  & & \mbox{CDF-N} & & \\\hline
GOODS-N ACS & 0.50 & 0.145 & 4.76\% & Closest-counterpart \\
GOODS-N IRAC & 0.75 & 0.202 & 3.99\% & Closest-counterpart \\
GALEX & 1.00 & 0.505 & 8.25\% & Closest-counterpart \\
GOODS-N $K_{\rm s}$ & 0.50 & 0.202 & 1.42\% & Closest-counterpart \\
GOODS-N MIPS & 0.75 & 0.276 & 6.05\% & Closest-counterpart \\
\mbox{2~Ms} \mbox{CDF-N} & 5.00 & 0.369 & 7.57\% & Likelihood-ratio\\ \hline
& & \mbox{CDF-S} & & \\\hline
MUSYC NIR & 0.75 & 0.167 & 3.53\% & Closest-counterpart \\
SIMPLE IRAC & 0.75 & 0.215 & 4.53\% & Closest-counterpart \\
GALEX & 1.00 & 0.475 & 8.95\% & Closest-counterpart \\
GOODS-S MIPS & 0.75 & 0.338 & 5.65\% & Closest-counterpart \\
\mbox{2~Ms} \mbox{CDF-S} & 5.00 & 0.302 & 7.04\% & Likelihood-ratio\\ \hline
\end{tabular}
\\All numbers quoted in this table are for matches to the base catalogs for sources with $m_{\rm R}\leq 26$ (only the central $r_{\rm encircled}=$8\arcmin \hspace{0.5mm} radius areas around the respective average aim points of the 2~Ms CDFs are considered; see \S~\ref{footprint} for source-selection areas).
Columns:
(1) Catalog that was matched to a base catalog. For the 2 Ms CDF catalogs, all \mbox{X-ray} sources (i.e., not limited to AGNs) were used for the assessments here.
(2) Maximum matching radius (for closest-counterpart matching) or searching radius (for likelihood-ratio matching).
(3) Median separation of all matches between a catalog and a base catalog.
(4) False-matching probability.
For the case of the 2 Ms CDF-N catalog, four filters ($U,R,I,HK^\prime$) of the North base catalog were used with likelihood-ratio matching.
The choice of these four filters was made to ensure broad wavelength coverage and to cover all the sources in the North base catalog using as few filters as possible (this criterion of choosing filters also applies to the South base catalog).
7.57\% is the average of the false rates derived with the four filters: 7.06\% for $U$-band, 8.66\% for $R$-band, 9.01\% for $I$-band, and 5.57\% for $HK^\prime$-band, respectively.
For the case of the 2 Ms CDF-S catalog, eight filters (MUSYC $B$, MUSYC $R$, MUSIC $I$, MUSIC $J$, COMBO-17 $U$, COMBO-17 $R$, COMBO-17 $915$ nm, and IRAC 3.6$\mu$m) of the South base catalog were used with likelihood-ratio matching. 7.04\% is the average of the false rates derived with the eight filters (in the aforementioned order, 8.61\%, 9.28\%, 6.83\%, 5.88\%, 7.31\%, 7.04\%, 7.31\%, and 4.05\%, respectively).
Note that the false rates from likelihood-ratio matching (i.e., 7.57\% and 7.04\%) not being smaller than those from closest-counterpart matching is due to the use of a larger maximum matching radius.
(5) Method that was used for matching catalogs.
\label{match}
\end{table}

The matching between the \mbox{X-ray} and optical/near-infrared sources is more challenging due to the fact that \mbox{X-ray} positions are usually not as good as the optical/near-infrared ones, which are 
accurate to \mbox{$\approx 0.1$--0.2\arcsec}\footnote{As inferred from Table~\ref{match}, \mbox{X-ray} positional accuracies are generally not as good as optical/IR ones, 
but are generally better than UV positions.
We note that these \mbox{X-ray} positional accuracies are the best so far at these fluxes due to the long exposures of the \mbox{2~Ms} CDFs:
the majority of sources within the central $r_{\rm encircled}=$8\arcmin \hspace{0.5mm} area have a positional accuracy of \mbox{$\lsim 0.4$--0.5\arcsec};
a number of sources even have a positional accuracy of \mbox{$\lsim 0.2$\arcsec} \hspace{0.35mm} that is comparable to the optical positional accuracy
(e.g., Alexander et al. 2003; Luo et al. 2008).};
the false-match probability increases because of the high density of faint background optical/near-infrared sources.
Therefore, we used likelihood-ratio matching (e.g., Ciliegi et al. 2003; Luo et al. 2010) between the \mbox{X-ray} and base catalogs, which not only takes into account the positional accuracy, but also the expected magnitude distribution of the counterparts. 
Likelihood-ratio matching outperforms closest-counterpart matching for the matching between the \mbox{X-ray} and optical sources, especially when matching faint sources (see Table~\ref{match}).\footnote{For the central $r_{\rm encircled}=$8\arcmin \hspace{0.5mm} area of the \mbox{CDF-N} (\mbox{CDF-S}), we obtained a false-match probability of 13.78\% (15.86\%) and a median offset of 0.364\arcsec \hspace{0.5mm} (0.291\arcsec) 
using closest-counterpart matching (compared to 7.57\% (7.04\%) and 0.369\arcsec \hspace{0.5mm} (0.302\arcsec) using likelihood-ratio matching; see Table~\ref{match}), with  
a matching radius of $r_{\rm m}=1.9\sqrt{\Delta_{\rm X}^2 + \Delta_{\rm O}^2}$ used, where 
$\Delta_{\rm X}$ ($\Delta_{\rm O}=0.1$\arcsec) is the \mbox{X-ray} (optical) positional error, and the coefficient of 1.9 was chosen to obtain roughly the same number of matches
as in the case of using likelihood-ratio matching for the purpose of direct comparison between methods. We refer readers to Luo et al. (2010) for more details on the likelihood-ratio matching method we used.}
We note that the same matching approach was used for both the \mbox{CDF-N} and \mbox{CDF-S}, with similar results (shown in Table~\ref{match}) produced for each field.

\subsection{Redshift Catalogs}\label{zcats}

We made use of both spectroscopic and photometric redshifts.
We preferred high-quality spectroscopic redshifts if available (see \S~\ref{zphot-quality} for more details); otherwise, we adopted photometric redshifts (Rafferty et al. 2010) derived with the Zurich Extragalactic Bayesian Redshift Analyzer 
(ZEBRA; Feldmann et al. 2006), using the ultradeep multiwavelength data described in \S~\ref{cats}. 
ZEBRA has features to correct systematic offsets in the photometry and construct new templates by modifying original templates based on the best fits between the photometry and original templates.
We produced $z_{\rm phot}$ catalogs for both the \mbox{CDF-N} (using data from up to 18 filters)
and \mbox{CDF-S} (using data from up to 42 filters). 
We note that our photometric redshifts have excellent quality considering dispersion and outliers
(see Rafferty et al. 2010; also see \S~\ref{zphot-quality}).
Luo et al. (2010) derived photometric redshifts for the 462 \mbox{X-ray} sources in the 2~Ms \mbox{CDF-S} main source catalog in a similar way, but with a more sophisticated treatment of photometry; manual source deblending was performed and upper limits were set in all photometric bands without detections.
Therefore, photometric redshifts from Luo et al. (2010) should on average have higher quality than our photometric redshifts and thus supersede our estimates if both estimates are available for a source (note that the two estimates are generally in good agreement with each other).  
We refer readers to Rafferty et al. (2010) and Luo et al. (2010) for the details on $z_{\rm phot}$ derivation and a complete list of $z_{\rm spec}$ references used.

\section{Source Properties}\label{properties}

\subsection{Rest-Frame Absolute Magnitudes}\label{absmag}

Using the data described in \S~\ref{cats} as well as upper limits in the $K$ and IRAC bands 
when appropriate,\footnote{For each of the $K$ and IRAC bands, a uniform upper limit is applied
if there is no detection at this specific band for a source. We find that
the utilization of upper limits significantly helps constrain the best-fit template for
sources with limited spectral coverage, and thus produces better estimates of physical properties (e.g., rest-frame absolute magnitudes, stellar masses, and SFRs) for these sources.\label{upperlimits}}
we computed rest-frame Johnson $U,B,V$ absolute magnitudes ($M_{\rm U},M_{\rm B},M_{\rm V}$) for each galaxy. 
We adopted the approach of template SED fitting to derive the absolute magnitudes, which was realized using ZEBRA. 
Compared to a linear (or log-linear) interpolation/extrapolation method, the template SED fitting reduces potential catastrophic failures,
especially in cases of limited/incomplete spectral coverage.   

First, we constructed a comprehensive set of galaxy and galaxy/AGN hybrid templates (see Rafferty et al. 2010 for details; also see Luo et al. 2010). Briefly, we used the 259 PEGASE galaxy templates that were employed by Grazian et al. (2006), which cover different galaxy types (elliptical, spiral, and starburst) and span a wide range of star-formation history and intrinsic extinction. We also used the 10 AGN templates of Polletta et al. (2007) that include a variety of empirical quasar and Seyfert templates. To account for both the nuclear and host-galaxy emission, we constructed a series of hybrid templates between the representative AGN templates and galaxy templates (e.g., Salvato et al. 2009). We chose 5 typical AGN templates (encompassing Type 1, Type 2, QSO 1, and QSO 2) and 16 typical PEGASE galaxy templates (encompassing elliptical, spiral, and starburst) for the construction of hybrid templates. For each pair of AGN and galaxy templates, the templates were normalized by the total integrated fluxes, and four hybrids with varying galaxy/AGN ratios (90:10, 75:25, 50:50, 25:75) were produced (note that the ratio of 10:90 is very similar to a 0:100 ratio, i.e., the original AGN templates, and thus was not included). Thus, we have a total of 330 galaxy/AGN hybrid templates (i.e., 330=$5\times 16\times 4+10$); the number of final templates is 259+330=589.

Second, we ran ZEBRA in Maximum-Likelihood mode to identify the best-fit template and compute absolute magnitudes for each source, with the input of photometry and redshift.
For moderate-luminosity AGNs that are best fit by hybrid templates, we used only galaxy-component SEDs to estimate their host physical properties (e.g., absolute magnitudes). The difference between these property estimates and those derived using hybrid template SEDs (i.e., galaxy-component plus AGN-component) is either negligible or small because the optical and near-infrared emission from moderate-luminosity active galaxies is dominated by host starlight (see \S~\ref{agn-contamination} for details).
Note that the absolute magnitudes output by ZEBRA are given in the AB system. In order to facilitate comparison with other works, these absolute AB magnitudes were converted into Vega magnitudes. 

We find our absolute magnitudes are in good agreement with those presented by Lehmer et al. (2008) for the overlapping sources, who adopted the approach of convolving photometrically derived SEDs with Johnson $U,B,V$ filter curves and directly computing rest-frame absolute magnitudes through interpolations and/or extrapolations, rather than our approach of template SED fitting.  
When compared to Lehmer et al. (2008), for sources in the \mbox{CDF-N} (\mbox{CDF-S}), the median ratio between the two sets of absolute magnitudes is $\approx 1.0$ ($\approx 1.0$) and there is 
a random scatter of $\lsim 0.2$ mag ($\lsim 0.3$ mag) [the median luminosity ratio is $\approx 1.0$ and random scatter is $\approx 0.1$ dex for both CDFs], which may be due to the fact that slightly different photometry and photometric redshifts were used.

\subsection{Stellar Masses}\label{stellarmass}

We estimated a set of stellar masses at various bands ($M_{\lambda,\star}$) for each source using the following tight correlations between rest-frame optical colors and stellar mass-to-light ratios,
\begin{equation}
{\rm log}(M_{\lambda,\star}/M_\odot)={\rm log}(L_\lambda/L_{\lambda,\odot})+b_\lambda(M_{\rm B}-M_{\rm V})+a_\lambda-0.10,
\label{ml_ratio}
\end{equation}
where $\lambda=B$, $V$, $R$, $I$, $J$, $H$, and $K$-band, respectively.
Equation~(\ref{ml_ratio}) was prescribed in Table~B1 of Zibetti, Charlot, \& Rix (2009) where the values of the coefficients ($a_\lambda$ and $b_\lambda$) can be found. 
The normalization in Eqn.~(\ref{ml_ratio}) has been adjusted by $-0.10$ dex to account for our adopted Kroupa (2001) initial mass function (IMF). 
We further used the relation $L_\lambda/L_{\lambda,\odot}=2.512^{M_{\lambda,\odot}-M_\lambda}$ for the mass estimation.
We note that Table~B1 of Zibetti, Charlot, \& Rix (2009) is a direct update of Table~7 of Bell et al. (2003) and
was derived using a technique that constructs spatially resolved maps of stellar-mass surface density in local galaxies based on optical and NIR imaging,
where the latest stellar population synthesis models were incorporated.
We chose to use Table~B1 of Zibetti, Charlot, \& Rix (2009) rather than Table~7 of Bell et al. (2003) because we find,
when using the prescription in Zibetti, Charlot, \& Rix (2009), that
(1) for the same color (e.g., $B-V$ in Eqn.~(\ref{ml_ratio})), the stellar-mass estimates derived at different bands are in better agreement with each other, in terms of both scaling and tightness; and
(2) for different colors (i.e., $B-V$ and $B-R$; the set of the coefficients, $a_\lambda$ and $b_\lambda$, for the $B-R$ color can also be found in Table~B1 of Zibetti, Charlot, \& Rix 2009)\footnote{We also estimated another set of stellar masses at various bands for each source using the $B-R$ color, again according to Table~B1 of Zibetti, Charlot, \& Rix (2009).
We find that the agreement between the two sets of stellar-mass estimates is good, so we 
adopted the $B-V$ color, i.e., Eqn.~(\ref{ml_ratio}), for the stellar-mass estimates, given the fact that the rest-frame $V$-band coverage is better than the
rest-frame $R$-band coverage for the sources.},
the stellar-mass estimates derived at the same band are in better agreement with each other, in terms of scaling.

Studies have shown that $K$-band galaxy luminosities are \mbox{5--10} times less sensitive to dust and stellar-population effects 
than optical luminosities (e.g., Bell \& de Jong 2000), which allows accurate stellar-mass estimations for galaxies.
Therefore, for each source, we consistently adopted the stellar mass derived with the reddest absolute magnitude as the final mass estimate, applying the condition that the source must actually have a corresponding rest-frame detection
(i.e., $M_{{\rm K},\star}$ has the highest priority and is adopted if a source has a rest-frame $K$-band detection).
Upper limits in the $K$ and IRAC bands used in the template SED fitting process (see \S~\ref{absmag}) were not counted for the purpose of determination of reddest rest-frame coverage, although they do help avoid unrealistically large mass estimates for the sources with limited spectral coverage.

For sources in the \mbox{CDF-S}, we compared our stellar-mass estimates with those presented by Borch et al. (2006) and find general agreement between methods, with a median ratio of $\approx 0.84$ between our and their estimates and \mbox{$\lsim$ 0.4} dex random scatter, which is adequate for our purposes in this work.\footnote{We note that estimates of stellar mass using photometry are generally accurate to within a factor of \mbox{$\approx 3$--5}.
We refer readers to other works for detailed discussion of the complexity and uncertainty of stellar-mass estimates (e.g., Bundy et al. 2006; Gallazzi \& Bell 2009; Muzzin et al. 2009; Zibetti, Charlot, \& Rix 2009).}
We speculate that the offset of mass scale may arise from the subtle differences in the IMF and the random scatter may mostly be due to the fact that we are using different photometric redshifts from those used by Borch et al. (2006).

\subsection{Star-Formation Rates}\label{sec_sfr}

We estimated the star-formation rate (SFR) of each source following Eqn. (1) in Bell et al. (2005):
\begin{equation}
{\rm SFR}(M_\odot\hspace{1mm}{\rm yr}^{-1})=9.8\times 10^{-11}(L_{\rm UV}/L_\odot+L_{\rm IR}/L_\odot),
\label{sfr_equ}
\end{equation}
where the solar bolometric luminosity $L_\odot=3.9\times 10^{33}$ \mbox{erg s$^{-1}$}. 
The dust-uncorrected ultraviolet luminosity $L_{\rm UV}$ was computed following $L_{\rm UV}=3.3\nu l_\nu(2800 \hspace{0.8mm}{\rm \AA})$ (see \S 3.2 of Bell et al. 2005), where $l_\nu(2800 \hspace{0.8mm}{\rm \AA})$ is the rest-frame 2800 $\AA$ monochromatic luminosity that was estimated using the best-fit template SEDs derived in \S~\ref{absmag}. 
For the estimation of the \mbox{8--1000 $\mu$m} infrared luminosity $L_{\rm IR}$, we employed an IDL routine ``chary\_elbaz\_24um.pro''\footnote{See http://david.elbaz3.free.fr/astro\_codes/chary\_elbaz.html.\label{ir-ext}} provided by Chary \& Elbaz (2001), which incorporates a library of 105 template SEDs that reproduce the observed 24 $\mu$m luminosity-$L_{\rm IR}$ correlations for local galaxies. 
We utilized observed MIPS \mbox{24 $\mu$m} fluxes (or upper limits) and redshifts to derive $L_{\rm IR}$ (or upper limits on $L_{\rm IR}$).
Papovich et al. (2007) found that SFRs derived with Eqn.~(\ref{sfr_equ}) (i.e., SFR$_{\rm UV+24 \mu m}$) are overestimated for luminous IR galaxies,
compared to the estimates of SFR$_{\rm UV+24, 70, 160 \mu m}$ that also consider longer-wavelength MIPS bands (i.e., 70 and 160 $\mu$m).
They corrected the overestimates using an empirical second-order polynomial that fits the correlation between log(SFR$_{\rm UV+24, 70, 160 \mu m}$) and log(SFR$_{\rm UV+24 \mu m}$).
We therefore applied the correction derived by Papovich et al. (2007) to our SFR$_{\rm UV+24 \mu m}$ estimates that are $\gsim 250$ $M_\odot$ yr$^{-1}$.
A few percent of the sources were corrected downward, with a multiplicative correction factor ranging from $\sim 0.3$--1.0 (the median correction factor is $\sim 0.8$).  

For sources with MIPS 24 $\mu$m detections in the \mbox{CDF-N} (\mbox{CDF-S}), we find good agreement between our SFR values and those derived by Lehmer et al. (2008), with a median ratio of $\approx 1.0$ ($\approx 1.0$) between our and their SFRs, and $\lsim 0.2$ ($\lsim 0.3$) dex random scatter. For sources without MIPS \mbox{24 $\mu$m} detections in the \mbox{CDF-N} (\mbox{CDF-S}), we find $\approx 100$\% ($\approx 100$\%) of our upper limits on SFR are consistent with those presented by Lehmer et al. (2008).
As a further check, we also computed SFR$_{\rm UV+70 \mu m}$ 
(i.e., using observed MIPS \mbox{70 $\mu$m} fluxes to derive $L_{\rm IR}$; an IDL routine ``chary\_elbaz.pro''$^{\ref{ir-ext}}$ was used)
for a subsample of 475 sources in the \mbox{CDF-S} base catalog that have both MIPS \mbox{24 $\mu$m} and \mbox{70 $\mu$m} detections, and compared with their corresponding SFR$_{\rm UV+24 \mu m}$ values.
We find good agreement between these two sets of SFR estimates;
the median ratio between SFR$_{\rm UV+70 \mu m}$ and SFR$_{\rm UV+24 \mu m}$ is $\approx 1.07$, with a random scatter of $\lsim 0.4$ dex and no systematic deviations.
This agreement further demonstrates that our SFR$_{\rm UV+24 \mu m}$ estimates are reliable and adequate for our purposes in this work.

\subsection{AGN Identification}\label{agn_id}

For the \mbox{2~Ms} CDF point-source catalogs, we used four primary criteria that rely upon distinct AGN physical properties to identify AGN candidates.
The techniques of AGN identification used here are detailed in, e.g., Bauer et al. (2004) and Lehmer et al. (2008).
Here, we only describe these techniques briefly. We note that a source can be identified as an AGN through multiple criteria.

(a) {\it \mbox{X-ray} Luminosity:}
The intrinsic \mbox{X-ray} luminosity ($L_{\rm X}$; quoted in the \mbox{0.5--8}~keV band throughout this paper) was estimated using the following equation:
\begin{equation}
L_{\rm 0.5-8\hspace{0.5mm}keV}=4\pi d_L^2 f_{\rm 0.5-8\hspace{0.5mm}keV,int}(1+z)^{\Gamma -2},
\end{equation}
where $d_L$ is the luminosity distance, $f_{\rm 0.5-8\hspace{0.5mm}keV,int}$ is the intrinsic (i.e., absorption-corrected) 
flux, and $\Gamma$ is the power-law photon index of the \mbox{X-ray} spectrum.
We derived $f_{\rm 0.5-8\hspace{0.5mm}keV,int}$ following these procedures:
(1) Using XSPEC (Arnaud 1996), we modeled the \mbox{X-ray} emission with an absorbed power-law model (both Galactic and intrinsic absorption),
which is defined as {\it zpow$\times$wabs$\times$zwabs} in XSPEC. 
The photon index was fixed to $\Gamma=1.8$ (typical for intrinsic AGN spectra) for each source and the redshifts of the {\it zpow} and {\it zwabs} 
components were fixed to that of the source. We additionally fixed the
Galactic column density to $N_{\rm H} = 1.6 \times 10^{20}$ cm$^{-2}$ for
the \mbox{CDF-N} and to $N_{\rm H} = 8.8 \times 10^{19}$ cm$^{-2}$ for
the \mbox{CDF-S} (Stark et al. 1992). 
(2) We used the above absorbed power-law model to find the intrinsic column density that reproduces the
observed band ratio, which is defined as the ratio of count rates between the hard (\mbox{2--8} keV) and soft (\mbox{0.5--2} keV) bands.
(3) We removed both Galactic and intrinsic absorption to obtain the intrinsic flux.
We find our $f_{\rm 0.5-8\hspace{0.5mm}keV,int}$ estimates to be in good agreement with those derived from 
spectral fitting utilizing ACIS Extract\footnote{See http://www.astro.psu.edu/xray/docs/TARA/ae\_users\_guide.html for details on ACIS Extract (Broos et al. 2010).} (F. Bauer et al., in preparation), in terms of correlation and dispersion.
 
Local \mbox{X-ray} observations show that purely star-forming non-AGN galaxies usually do not have intrinsic $L_{\rm X}\geq 3\times 10^{42}$ \mbox{erg s$^{-1}$}. 
We thus classify a source with $L_{\rm X}\geq 3\times 10^{42}$ \mbox{erg s$^{-1}$} as an AGN (i.e., a luminous AGN).
We note that there could be potential intruders such as high-redshift strongly star-forming sources (e.g., submillimeter galaxies) that
have a substantial amount of \mbox{X-ray} emission mainly from active stellar populations (e.g., Alexander et al. 2005a).
However, the $L_{\rm X}$-to-SFR ratio of these star-forming galaxies 
is typically lower than that of AGNs (see Criterion (c) below),
which can be used as a discriminator to differentiate these non-AGN galaxies from AGNs.
Indeed, we find that over $\approx 92$\% of the 2~Ms CDF AGNs identified with Criterion (a) were also identified with the $L_{\rm X}$-to-SFR ratio criterion (i.e.,
Criterion (c)), which suggests that contamination by high-redshift star-forming galaxies with Criterion (a) is not significant in our case.     

(b) {\it \mbox{X-ray} Hardness:}
One signature of moderately-to-highly obscured ($N_{\rm H}\gsim 10^{22}$ cm$^{-2}$) AGNs is their hard \mbox{X-ray} spectra.
If a source has an effective photon index of $\Gamma_{\rm eff} \leq 1.0$, we flag it as an AGN (i.e., an obscured AGN).

(c) {\it X-ray-to-SFR Correlation:}
AGNs often show significant \mbox{X-ray} luminosity excesses over what is expected based on the \mbox{$L_{\rm X}$-SFR} correlation. 
We modified the $L_{\rm X}$-SFR correlation shown in Persic \& Rephaeli (2007) to $L_{\rm 0.5-8\hspace{0.5mm}keV}/{\rm SFR}=1.06\times 10^{40}$ erg s$^{-1}$ ($M_\odot$ yr$^{-1}$)$^{-1}$,
taking into account the fact that we computed SFRs using Eqn.~(\ref{sfr_equ}).
If a source has $L_{\rm 0.5-8\hspace{0.5mm}keV}/{\rm SFR}\gsim 3$ times that from the above $L_{\rm X}$-SFR correlation, which corresponds to a factor of $\gsim 2.5$ times the RMS scatter of the Persic \& Rephaeli $L_{\rm X}$-SFR correlation, we classify it as an AGN.

(d) {\it X-ray-to-Optical Flux Ratio:} 
The X-ray-to-optical flux ratio ${\rm log}(f_{\rm 0.5-8\hspace{0.5mm}keV,int}/f_{\rm R})=-1$ is regarded as a useful AGN/galaxy discriminator.
We mark a source as an AGN if it satisfies ${\rm log}(f_{\rm 0.5-8\hspace{0.5mm}keV,int}/f_{\rm R})>-1$.

\section{Sample Construction}\label{sec:sample}

We constructed a parent sample within the \mbox{2~Ms} CDFs that consists of 207 X-ray selected moderate-luminosity AGNs and a population of 12,329 galaxies that the AGNs were drawn from.
We then extracted four samples, including both stellar-mass and luminosity selected samples, from this parent sample for various purposes.
The details of sample construction are described below.
 
\subsection{Source-Selection Areas}\label{footprint}

We restricted our source selection to the central $r_{\rm encircled}=$8\arcmin \hspace{0.5mm} radius areas around the respective average aim points of the \mbox{2~Ms} CDFs that are weighted by exposure time
(for the \mbox{CDF-N}: $\alpha_{\rm J2000.0}=12^{\rm h}36^{\rm m}45.^{\rm s}7$, $\delta_{\rm J2000.0}=+62^{\rm o} 13\arcmin 58\arcsec$; 
for the \mbox{CDF-S}: $\alpha_{\rm J2000.0}=03^{\rm h}32^{\rm m}28.^{\rm s}8$, $\delta_{\rm J2000.0}=-27^{\rm o} 48\arcmin 23\arcsec$),
and we also required GOODS-N/GOODS-S coverage with {\it HST} of sources (only $\approx$ 15\% of the central $r_{\rm encircled}=$8\arcmin \hspace{0.5mm} areas do not have GOODS coverage).
The corresponding total selected area is $\approx$ 314 arcmin$^2$ (\mbox{$\approx$ 155} arcmin$^2$ for the \mbox{CDF-N} and \mbox{$\approx$ 159} arcmin$^2$ for the \mbox{CDF-S}).
The reasons for this areal choice are
(1) The central CDF areas provide the deepest \mbox{X-ray} coverage to date, with sensitivity limits of \mbox{$\lsim 2.3\times 10^{-16}$} \mbox{erg cm$^{-2}$ s$^{-1}$} for the \mbox{0.5--8}~keV band. Such deep \mbox{X-ray} data can reduce the bias that shallower \mbox{X-ray} surveys have against finding AGNs in less-massive hosts (e.g., Merloni \& Heinz 2008). \mbox{X-ray} surveys have been extremely efficient in defining reliable AGN samples; the \mbox{2~Ms} CDFs have reached an AGN density of \mbox{$\approx 7200$} deg$^{-2}$ (e.g., Bauer et al. 2004).
(2) More accurate \mbox{X-ray} positions for sources in the central areas lead to more reliable matching between \mbox{X-ray} and faint optical sources (see \S~\ref{matching}).
(3) The redshift success rate (i.e., availability of spectroscopic redshifts, $z_{\rm spec}$) is relatively high due to intensive spectroscopic investment in these areas.
Furthermore, the quality of photometric redshifts ($z_{\rm phot}$) has been greatly improved utilizing the superb multiwavelength data available.
(4) The deep multiwavelength coverage within the GOODS areas (e.g., ultradeep {\it HST} observations provide the best optical photometry and images for analysis) allows for the construction of well-sampled rest-frame spectral energy distributions (SEDs) that are key to estimations of $z_{\rm phot}$ and physical properties.

For the \mbox{CDF-N}, there are 11,951 galaxies (253 AGNs) located within the northern $r_{\rm encircled}=$8\arcmin \hspace{0.5mm} area and having GOODS coverage;
for the \mbox{CDF-S}, there are 21,749 galaxies (250 AGNs) located within the southern $r_{\rm encircled}=$8\arcmin \hspace{0.5mm} area and having GOODS coverage.\footnote{Note that the total galaxy number
in the respective central $r_{\rm encircled}=$8\arcmin \hspace{0.5mm} area and having GOODS coverage significantly differs between the North and South base catalogs, i.e., 11,951 vs. 21,749,
which is mainly due to the fact that  
the South base catalog was constructed in a different way from the North, and it includes a large number of faint sources (e.g., $m_{\rm R}>26$) because the GOODS-S MUSIC catalog has a set of fainter optical and near-infrared detection limits than the North base catalog.}

\subsection{Filter-Number Cut}\label{nfilter}

To ensure the reliability of the observed source SEDs, we discarded sources that have detections in less than five filters.
Consequently, the above galaxy numbers of 11,951 and 21,749 are reduced to 11,949 (\mbox{$\approx 0.02$\%} loss) and 21,376 (\mbox{$\approx 1.72$\%} loss), respectively.
We do not expect this filter-number cut to introduce any significant bias because only very small fractions of sources were dropped, and dropped sources are typically very faint (and will be excluded later anyway; see the \mbox{$m_{\rm R}\leq 26$} cut in \S~\ref{cuts}).
We note that the numbers of AGNs in the \mbox{CDF-N} and \mbox{CDF-S} are not affected by this cut (still 253 and 250, respectively).  

\subsection{Redshift Cut and Magnitude Cut}\label{cuts}

In addition to our areal choice and filter-number cut, we limited our sample selection to a redshift range of \mbox{$0<z\leq 4$}.
We further applied an \mbox{$R$-band} magnitude cut of \mbox{$m_{\rm R}\leq 26$} for the source selection, which ensures that the sources in our sample have high-quality photometry and photometric redshifts, and are only subject to minimal cross-matching failures.
Complexity in the application of a uniform cut of \mbox{$m_{\rm R}\leq 26$} arose from the fact that different \mbox{$R$-band} filters were used for the \mbox{CDF-N} (the Subaru $R$ broad-band filter) 
and \mbox{CDF-S} (the COMBO-17 and MUSYC $R$ broad-band filters) observations. 
We converted the two \mbox{CDF-S} \mbox{$R$-band} magnitudes to the Subaru \mbox{$R$-band} magnitude using a \mbox{$K$-correction} package (kcorrect.v4\_1\_4\footnote{See http://cosmo.nyu.edu/blanton/kcorrect/.}; Blanton \& Roweis 2007), 
which convolves the photometrically derived SED (see \S~\ref{cats} for the photometry catalogs used) of a source with the above three $R$-band filter curves and computes the difference ($\Delta m_{\rm R}$) between the derived $R$-band magnitudes. 
Typically, $\Delta m_{\rm R}$ is small ($|\Delta m_{\rm R}|\lsim 0.15$) with a median value of $\approx 0.02$. 
After applying the redshift and magnitude cuts, our sample consists of 401 AGNs and 19,202 galaxies. 

\subsection{Stellar-Mass Cut and/or Luminosity Cut}\label{ml_cuts}

\begin{figure}[ht]
\includegraphics[width=3.3in]{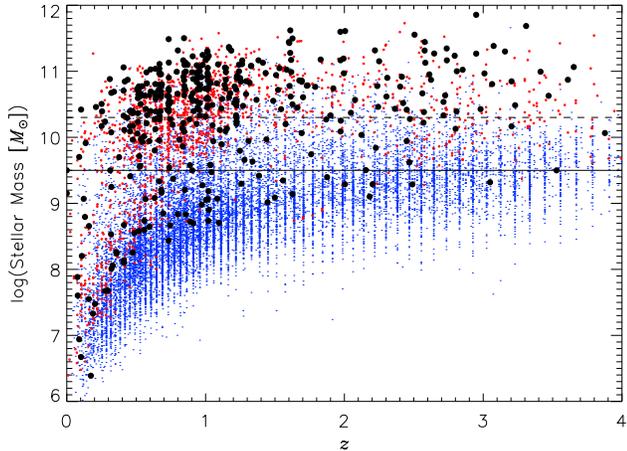}
\caption{Stellar mass as a function of redshift for the sample that consists of 401 AGNs and their 19,202 parent galaxies (see \S~\ref{cuts}). AGNs are shown as black large filled circles; red/blue galaxies are shown as red circles/blue dots (red and blue galaxies are separated using Eqn.~(\ref{separation-line}) in \S~\ref{sec:bi-gal}). The solid (dashed) line indicates the stellar-mass cut of $10^{9.5}M_\odot$ ($10^{10.3}M_\odot$), above which blue (red) galaxies in \mbox{Sample A} are roughly complete (see \S~\ref{sample}).
The apparent redshift quantization reflects the logarithmic redshift steps
we adopted for deriving the photometric redshifts. (A color version of this figure is available in the online journal.)}
\label{mass-z}
\end{figure}

Stellar mass is likely the most fundamental observable parameter for understanding the properties of galaxies.
Using stellar-mass selected samples to study galaxy evolution has some advantages over using either color- or luminosity-selected samples, e.g., avoiding biases associated with sample selections (see \S~\ref{summary} for details).
Figure~\ref{mass-z} shows stellar mass as a function of redshift for our sample of 401 AGNs and their 19,202 parent galaxies (see \S~\ref{cuts}).
According to Fig.~\ref{mass-z}, our sample of blue (red) galaxies is roughly complete above $10^{9.5}M_\odot$ ($10^{10.3}M_\odot$) from \mbox{$z\approx 0$} to \mbox{$z\approx 4$}
(red and blue galaxies are separated using Eqn.~(\ref{separation-line}) in \S~\ref{sec:bi-gal}).
The completeness limit of $10^{9.5}M_\odot$ ($10^{10.3}M_\odot$) is confirmed using the technique presented in \S~5.1 of Meneux et al. (2008)
that constructs the mass completeness limit as a function of redshift based on the properties (i.e., rest-frame absolute magnitude, redshift, and stellar mass) of galaxies in a sample. 
We therefore applied a stellar-mass cut of \mbox{$\geq 10^{9.5}M_\odot$} to our sample.

To make direct comparisons with previous CMD studies that typically made use of luminosity limited samples, 
we also kept sources that satisfy a luminosity cut of $M_{\rm V}\leq -19$
(i.e., we included both stellar-mass and luminosity selected samples in our parent sample; see \S~\ref{sample_def}).
Therefore, there are 12,329 galaxies (375 AGNs) in our sample, which have
either $M_\star\geq 10^{9.5}M_\odot$ or $M_{\rm V}\leq -19$. 

\subsection{X-ray Luminosity Cut and Removal of Broad-Line AGNs}\label{lum-cut}

We further chose to restrict the luminosity range of AGNs to $41.9\leq {\rm log}(L_{\rm X}/({\rm erg\:s^{-1}}))\leq 43.7$ (Silverman et al. 2008b).
The choice of the lower and upper luminosity limits was made to remove the contamination of any non-AGN galaxies having significant \mbox{X-ray} emission,
to minimize potential luminosity-dependent effects,
and to ensure that the optical and near-infrared emission is primarily due to host galaxies (see \S~2.2 of Silverman et al. 2008b for details; also see \S~\ref{agn-contamination}).
There are 215 AGNs (out of 375; see \S~\ref{ml_cuts}) that satisfy this \mbox{X-ray} luminosity cut:
$215=375-89-71$, where 89 AGNs have ${\rm log}(L_{\rm X}/({\rm erg\:s^{-1}}))<41.9$ and 71 AGNs have ${\rm log}(L_{\rm X}/({\rm erg\:s^{-1}}))>43.7$.
We note that all of the 89 AGNs with ${\rm log}(L_{\rm X}/({\rm erg\:s^{-1}}))<41.9$ dropped have $z\lsim 1.0$, which is not the redshift range of primary interest for this study.
Furthermore, we have verified that our basic results do not change with these 89 AGNs included. 

Even with all the above source-filtering procedures, a few broad-line AGNs (BLAGNs) can still survive in our sample that are optically bright and
significantly contribute to the optical emission; we cannot obtain reliable stellar-mass constraints for these sources.
We thus used the identifications of BLAGNs in the CDFs (Barger et al. 2003; Szololy et al. 2004) to remove the BLAGNs remaining in our sample.
After this screening, there are 207 AGNs remaining in our sample (i.e., 8 BLAGNs were dropped). 

\subsection{Sample Properties}\label{sample}

\subsubsection{Sample Definitions}\label{sample_def}

Using the above criteria ($r_{\rm encircled}=8$\arcmin, having GOODS coverage, having no less than 5 filters, $0<z\leq 4$, $m_{\rm R}\leq 26$, $M_\star\geq 10^{9.5}M_\odot$ or $M_{\rm V}\leq -19$, $41.9\leq {\rm log}(L_{\rm X}/({\rm erg\:s^{-1}}))\leq 43.7$, and not being a BLAGN; the latter two criteria are applied only to AGNs), we constructed a parent sample that consists of 207 \mbox{X-ray} selected AGNs, which was drawn from a population of 12,329 galaxies. 
Table~\ref{sample_const} shows a summary of parent sample construction.

\begin{table}[ht]
\caption{Summary of Parent Sample Construction}
\begin{tabular}{lrrrr}\hline\hline
         & Galaxies & Galaxies & AGNs & AGNs \\
Criteria & (CDF-N)  & (CDF-S)  & (CDF-N) & (CDF-S) \\\hline
1. $r_{\rm encircled}=8$\arcmin, GOODS coverage & 11,951 &  21,749  &    253   &   250\\
2. $\geq 5$ filters &      11,949 & 21,376 &     253  &    250\\
3. $0<z\leq 4$, $m_{\rm R}\leq 26$ &       8588 & 10,614  &    230  &    171\\
4. $M_\star\geq 10^{9.5} M_\odot$ or $M_{\rm V}\leq -19$ & 5738 & 6591 & 214 & 161 \\
5. $10^{41.9}\leq L_{\rm X}\leq 10^{43.7}$ &      --- & --- &      116  &     99\\
6. Not a BLAGN &  --- & --- &  115       &  92  \\\hline
Parent Sample &  5738 & 6591  & 115   &  92  \\\hline
\end{tabular}
\label{sample_const}
\end{table}

We extracted four samples (A, B, C, and D) from the parent sample that are appropriate for various purposes.
Table~\ref{samples} shows some basic information about these four samples.
In \mbox{Sample A} ($0<z\leq 4$ and $M_\star\geq 10^{9.5}M_\odot$), blue galaxies with $M_\star\geq 10^{9.5}M_\odot$ are roughly complete from \mbox{$z\approx 0$--4}, whereas the completeness of red galaxies is not as high (see Fig.~\ref{mass-z}).
We used \mbox{Sample A} mainly for the examination of possible galaxy color bimodality up to high redshifts (see \S~\ref{sec:bi}) because \mbox{Sample A}, consisting of a sufficient number of blue and red galaxies, is suitable for this purpose.
There are only a few AGNs with $3<z\leq 4$ in \mbox{Sample A}.
We thus focused on a redshift subrange of $0<z\leq 3$ in \mbox{Sample A} for studying both AGNs and their hosts, using \mbox{Sample B} ($0<z\leq 3$ and $M_\star\geq 10^{9.5}M_\odot$) and \mbox{Sample C} ($0<z\leq 3$ and $M_\star\geq 10^{10.3}M_\odot$) that are two subsets of \mbox{Sample A}.
We note that the mass cut of $10^{10.3} M_\odot$ in \mbox{Sample C} is the mass above which both AGN hosts and non-AGN galaxies have similar distributions of rest-frame $U-V$ color (see \S~\ref{color-dep}).
In \mbox{Sample B}, only blue galaxies are roughly complete;
while in \mbox{Sample C}, both blue and red galaxies are roughly complete (see Fig.~\ref{mass-z}).
Therefore, these two samples were often used together to examine the effects of sample incompleteness on various results (see \S~\ref{sec:cmd} and \ref{color-dep}).
\mbox{Sample D} ($0<z\leq 3$ and $M_{\rm V}\leq -19$) was used to represent luminosity limited samples that have typically been adopted in previous CMD studies; 
the CMD results obtained with \mbox{Sample D} were compared with those from mass-limited samples (i.e, Samples A, B, and C, which are the main focus of this work).
In the remainder of \S~\ref{sample}, we describe the properties of \mbox{Sample A}, which includes both \mbox{Sample B} and \mbox{Sample C}.

\begin{table*}[ht]
\caption{Galaxy and AGN Samples Drawn from Parent Sample}
\resizebox{\textwidth}{!}{%
\begin{tabular}{ccccc}\hline\hline
Sample & Condition & No. of Galaxies & No. of AGNs & Main purpose \\\hline
Sample A & $0<z\leq 4$, $M_\star\geq 10^{9.5}M_\odot$ & 4941 & 188 & Color bimodality \\
Sample B & $0<z\leq 3$, $M_\star\geq 10^{9.5}M_\odot$ & 4357 & 185 & CMD, color dependence, AGN fraction, etc. \\
Sample C & $0<z\leq 3$, $M_\star\geq 10^{10.3}M_\odot$ & 1468 & 139 & CMD, color dependence, AGN fraction, etc. \\
Sample D & $0<z\leq 3$, $M_{\rm V}\leq -19$ & 11,119 & 202 & CMD, color dependence, AGN fraction, etc. \\\hline
\end{tabular}}
\label{samples}
\end{table*}

\subsubsection{Quality of Photometric Redshifts}\label{zphot-quality}

\begin{figure*}[ht]
\includegraphics[width=7.1in]{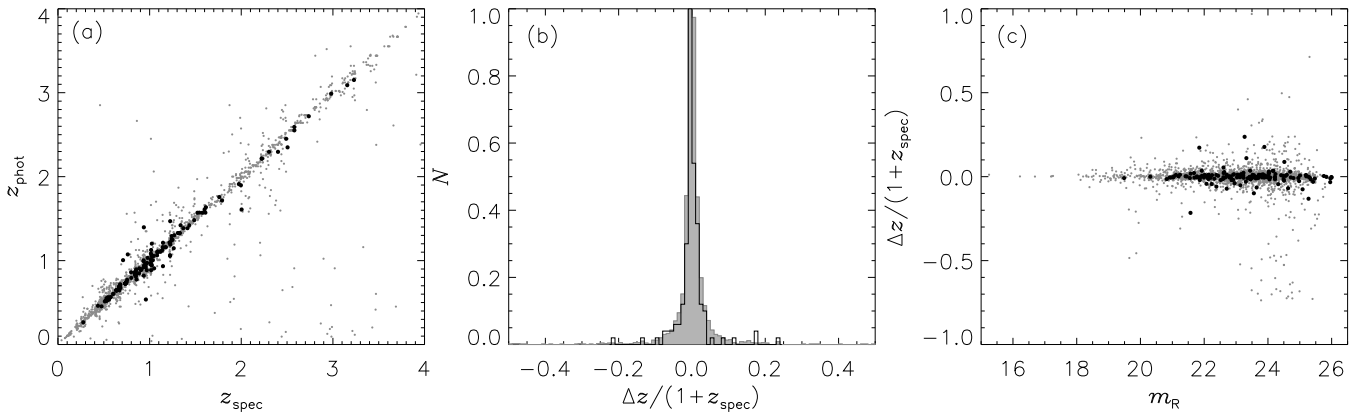}
\caption{Plots of $z_{\rm phot}$ quality checks for \mbox{Sample A}, with AGNs shown as filled circles/solid histogram and non-AGN galaxies shown as grey dots/grey histogram.
(a) $z_{\rm phot}$ vs. $z_{\rm spec}$.
(b) Normalized histograms of $\Delta z/(1+z_{\rm spec})$.
The histograms largely overlap each other because AGNs and non-AGN galaxies have photometric redshifts of similar quality.
(c) $\Delta z/(1+z_{\rm spec})$ as a function of $R$-band magnitude $m_{\rm R}$.}
\label{zphot}
\end{figure*}

We assessed the $z_{\rm phot}$ quality for \mbox{Sample A} (see Fig.~\ref{zphot}) by using a number of quantities produced by a comparison of the photometric and spectroscopic redshifts of the sample: the normalized median absolute deviation ($\sigma_{\rm NMAD}=1.48\times {\rm median}[|(\Delta z - {\rm median}(\Delta z))/(1+z_{\rm spec})|]$), the average absolute scatter ($AAS={\rm mean} [|\Delta z/(1+z_{\rm spec})|]$), and the percentages of outliers ($P_{\rm 0.2(0.1)\_outlier}$) with $|\Delta z|/(1+z_{\rm spec}) > 0.2(0.1)$, where $\Delta z=z_{\rm phot}-z_{\rm spec}$. For the AGNs, we find that $\sigma_{\rm NMAD}=0.0154$, $AAS=0.0214$, $P_{\rm 0.2\_outlier}=1.35$\%, and $P_{\rm 0.1\_outlier}=4.05$\%; for the galaxies, we find that $\sigma_{\rm NMAD}=0.0153$, $AAS=0.0323$, $P_{\rm 0.2\_outlier}=2.54$\%, and $P_{\rm 0.1\_outlier}=5.51$\%.
We owe such a good quality of $z_{\rm phot}$ to the large CDF multiwavelength observational investments and continuous efforts of improving $z_{\rm phot}$ estimation techniques (see, e.g., Luo et al. 2010; Rafferty et al. 2010; and references therein). 
However, as cautioned by Luo et al. (2010) and Rafferty et al. (2010), 
the spectroscopic subsample is likely to have significantly better photometric redshifts than the full sample due to two reasons:
(1) the sources in the spectroscopic subsample are generally brighter and therefore not entirely representative of the full sample;
and (2) the template-improvement step (see \S~\ref{zcats}) used in the $z_{\rm phot}$ derivation
optimizes the templates for the spectroscopic subsample and thus introduces a bias that is favorable to 
the spectroscopic subsample.\footnote{In order to assess the actual quality of our photometric redshifts,
we repeated a divided-sample test several times: 
we ran template-improvement mode with $\approx 3/4$ of the sources in the spectroscopic subsample that were randomly selected
and then applied the obtained improved templates to the unselected $\approx 1/4$ of the sources to test for quality. 
The tests suggest that the percentage of outliers for the non-trained sources is larger than that for the trained sources:
$P_{\rm 0.2\_outlier}\approx 6.5$\% ($\approx 14.9$\%) for the non-trained galaxies (AGNs).}

We do not expect the photometric redshift uncertainties to introduce a systematic bias in the estimation of source properties (e.g., rest-frame absolute magnitudes) 
because there is no systematic offset between photometric redshifts and spectroscopic redshifts and the residuals between photometric redshifts and spectroscopic redshifts are approximately symmetric (see Fig.~\ref{zphot}).
Table~\ref{zspec_percent} shows $z_{\rm spec}$ fractions of the sources within different redshift ranges in \mbox{Sample A}, which indicates that the $z_{\rm spec}$ fractions for AGNs are quite high and those for galaxies are reasonably high, especially at low redshifts.

\begin{table}[ht]
\caption{$z_{\rm spec}$ Fractions for Different Redshift Bins in Sample A}
\begin{tabular}{ccc}\hline\hline
Redshift & No. of Galaxies (\% of $z_{\rm spec}$) & No. of AGNs (\% of $z_{\rm spec}$) \\\hline
$0<z\leq 1$ & 1626 (68.7\%) & 83 (94.0\%) \\
$1<z\leq 2$ & 1570 (36.5\%) & 85 (71.8\%)  \\ 
$2<z\leq 3$ & 1161 (15.0\%) & 17 (47.1\%) \\
$3<z\leq 4$ &  584 (10.6\%) & 3  (33.3\%) \\
$0<z\leq 3$ & 4357 (42.8\%) & 185 (79.5\%) \\
$0<z\leq 4$ & 4941 (39.0\%) & 188 (78.7\%) \\\hline
\end{tabular}
\label{zspec_percent}
\end{table}

\subsubsection{AGN Contamination}\label{agn-contamination}

\begin{figure*}[ht]
\includegraphics[width=7.1in]{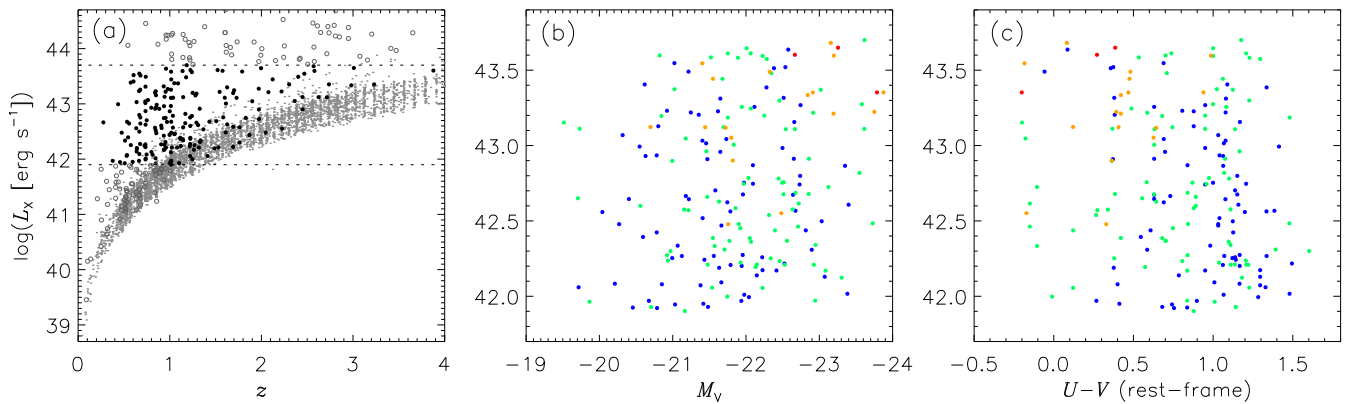}
\caption{(a) 0.5--8 keV luminosity as a function of redshift for \mbox{Sample A} (see \S~\ref{sample}).
Filled circles indicate AGNs and small grey dots indicate non-AGN galaxies with their upper limits derived from corresponding sensitivity maps.
The spread of a factor of $\approx 3$ in the upper limits at a given redshift
reflects the sensitivity variation with off-axis angle, i.e., a smaller off-axis angle corresponds to a higher sensitivity.
The two dashed lines show the \mbox{X-ray} luminosity cut of $41.9\leq {\rm log}(L_{\rm X}/({\rm erg\:s^{-1}}))\leq 43.7$.
For comparison, AGNs that lie outside the luminosity range of $41.9\leq {\rm log}(L_{\rm X}/({\rm erg\:s^{-1}}))\leq 43.7$ are shown as grey circles.
(b) \mbox{0.5--8}~keV luminosity as a function of $M_{\rm V}$ for the AGNs in \mbox{Sample A}.
(c) \mbox{0.5--8}~keV luminosity as a function of rest-frame $U-V$ color for the AGNs in \mbox{Sample A}.
AGNs in (b) and (c) are color-coded such that the blue (green, orange, red) color represents $0<z\leq 1$ ($1<z\leq 2$, $2<z\leq 3$, $3<z\leq 4$).
No apparent correlation is seen in (b) or (c), demonstrating that the optical emission of these AGN hosts should not be dominated by the active nuclei. (A color version of this figure is available in the online journal.)}
\label{lum-z}
\end{figure*}

\begin{figure}[ht]
\includegraphics[width=3.3in]{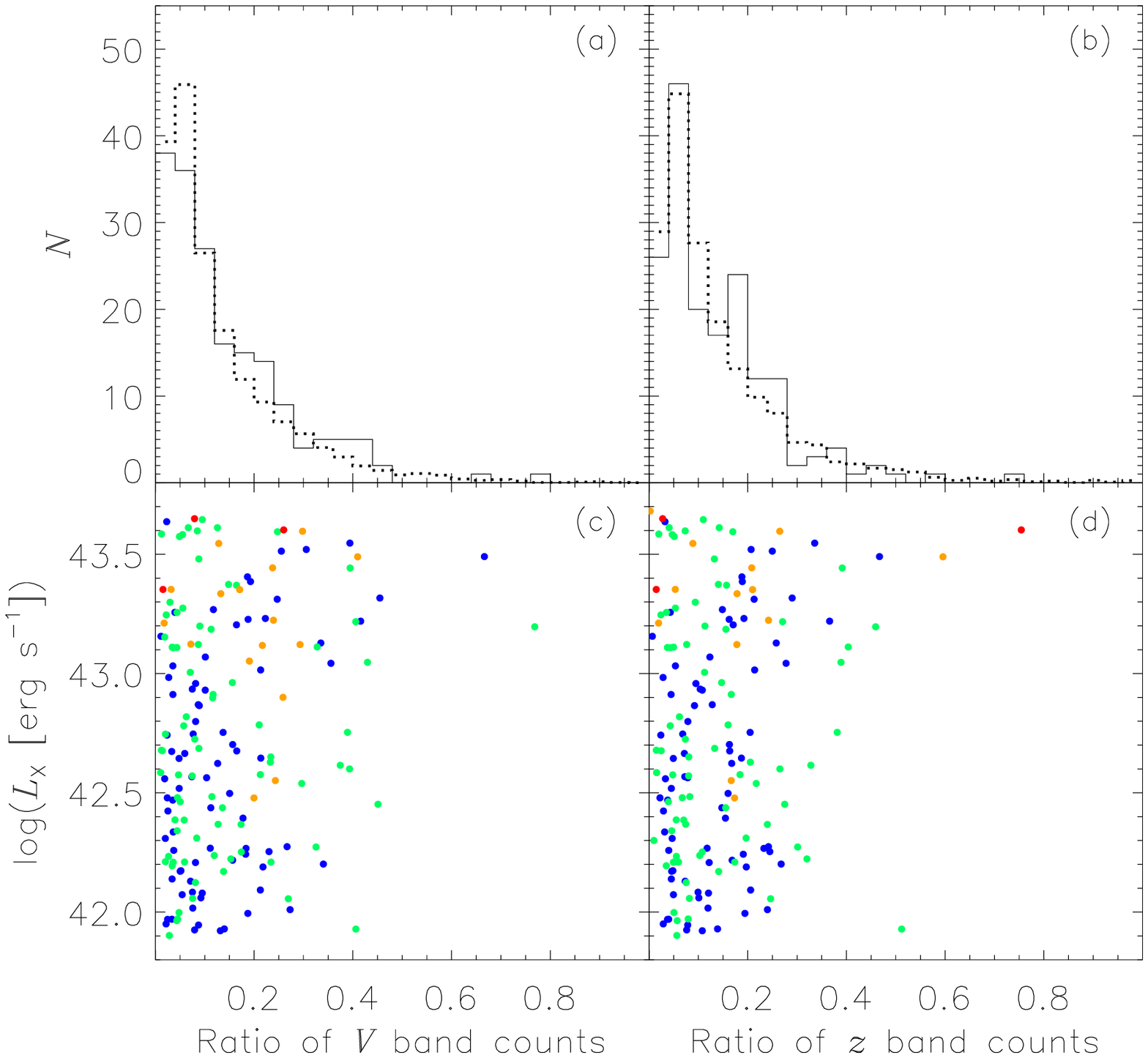}
\caption{AGN contribution to the total (galaxy+AGN) optical emission in the $V_{606}$-band and $z_{850}$-band (see \S~\ref{lum-cut}) for \mbox{Sample A}.
The ratio of counts is computed by measuring optical counts 
in circular apertures of
two different sizes ($r=3$ and 25 pixels, i.e., $r=0.09$\arcsec \hspace{0.3mm} and 0.75\arcsec, respectively) located at the centroid of the optical emission.
These ratios should provide a firm upper limit to the AGN contribution since we did not remove stellar emission within the small aperture.
(a, b) Histograms of the ratio of counts of the AGN hosts (solid histograms) and non-AGN galaxies (dotted histograms),
where the non-AGN galaxy numbers have been scaled to match those of the AGN hosts.
The majority of the AGN hosts have $\gsim 75$\% of their total optical emission (in the $V_{606}$ and $z_{850}$ bands) from outside of $r=0.09$\arcsec \hspace{0.2mm}
and have count ratios similar to the parent population of galaxies.
(c, d) \mbox{X-ray} luminosity vs. ratio of counts for the AGN hosts in the $V_{606}$-band and $z_{850}$-band. There is no clear correlation between \mbox{X-ray} luminosity and the ratio of counts for the AGN hosts, which supports the conclusion that the optical emission is not dominated by the AGN component.
We note that there are a few AGN hosts that have a ratio of counts greater than 0.5,
which should not be a problem because their
optical images appear normal and the ratio of counts only represents an upper limit of AGN contribution to the total optical emission.
AGNs in (c) and (d) are color-coded in the same way as those in Figs.~\ref{lum-z}(b, c). (A color version of this figure is available in the online journal.)}
\label{ratio}
\end{figure}

The X-ray luminosity cut of $41.9\leq {\rm log}(L_{\rm X})\leq 43.7$ (see \S~\ref{lum-cut}) was adopted to minimize potential AGN contamination of optical and near-infrared emission from host galaxies.
Below we present four additional arguments to show that the optical and near-infrared emission from the 188 active galaxies in \mbox{Sample A} is dominated by host starlight and not strongly influenced by the AGNs.
(1) Of these 188 AGNs, 186 ($\approx 99$\%) have best-fit SED templates with the optical and near-infrared emission dominated by galaxy starlight.
(2) Figure~\ref{lum-z} shows \mbox{0.5--8}~keV luminosities of these 188 AGNs as a function of redshift, \mbox{$V$-band} absolute magnitude,
and rest-frame $U-V$ color (i.e., $M_{\rm U}-M_{\rm V}$), respectively. According to Figs.~\ref{lum-z}(b, c), there is no clear correlation between \mbox{X-ray}
luminosity and either $M_{\rm V}$ or $U-V$.
This is consistent with the \mbox{X-ray} and optical emission not being produced by the same process, suggesting that the optical emission is not likely to be dominated by the AGN component.
(3) Following Silverman et al. (2008), we determined conservative upper limits on the AGN contribution to the total (galaxy+AGN) optical emission 
using the {\it HST} ACS $V_{606}$- and $z_{850}$-band images (rescaled to 0.03\arcsec \hspace{0.3mm} pixel$^{-1}$) provided by GOODS.  
About a dozen AGNs were not included in the analysis due to their proximity to the ACS field edge.
We measured optical counts 
in circular apertures of
two different sizes ($r=3$ and 25 pixels, i.e., $r=0.09$\arcsec \hspace{0.3mm} and 0.75\arcsec, respectively) located at the centroid of the optical emission.
$r=0.09$\arcsec \hspace{0.3mm} ($r=0.75$\arcsec) covers a physical scale of \mbox{0.378--0.763} \mbox{(3.147--6.354) kpc} for these AGNs with \mbox{$z\approx 0.276$--3.785}.
Figures~\ref{ratio}(a) and \ref{ratio}(b) show the histograms of the ratio of counts between these two apertures for the AGN hosts for the two ACS bands, respectively.
We note that these ratios provide a firm upper limit to the AGN contribution since we did not remove stellar emission within the small aperture.
For the $V_{606}$-band ($z_{850}$-band), the mean ratio of counts of these AGNs is 0.142 (0.139), and 84.2\% (87.2\%) of them have a ratio less than 0.25,
which indicate that the optical emission is not dominated by the AGN component.
Note also that these AGN hosts have count ratios similar to the parent population of galaxies.
(4) According to Figures~\ref{ratio}(c) and \ref{ratio}(d), there is no apparent correlation between \mbox{X-ray} luminosity and 
the ratio of counts for the AGN hosts, which supports the conclusion that the optical emission is not dominated by the AGN component.
Generally speaking, galaxy surface brightness dims with increasing redshift and 
the contrast ratio between a nuclear point source and its  
host galaxy increases toward higher redshifts.
This could potentially impact our conclusion (for high-redshift AGNs) 
that the optical emission is not dominated by the AGN component.
However, as shown in Figs.~\ref{ratio}(c, d), this does not appear to be an issue
with AGNs in \mbox{Sample A} because the ratio of counts does not tend to
be larger for sources with larger redshifts.

\begin{figure}[ht]
\includegraphics[width=3.3in]{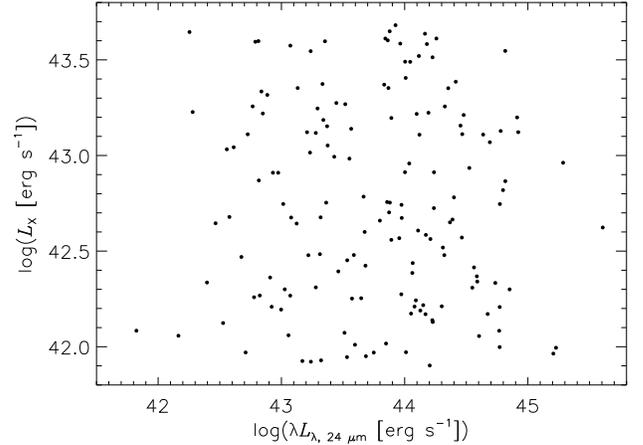}
\caption{X-ray luminosity as a function of MIPS $\lambda L_{\lambda, 24 \mu \rm m}$ for the 157 AGNs in \mbox{Sample A} that have MIPS 24 $\mu$m detections. There is no apparent correlation between $L_{\rm X}$ and $\lambda L_{\lambda, 24 \mu \rm m}$, which suggests that these AGNs do not significantly affect the observed \mbox{$24 \mu m$} fluxes that are used to compute the SFRs.}
\label{lx-mips}
\end{figure}

Figure~\ref{lx-mips} shows \mbox{X-ray} luminosity as a function of MIPS $\lambda L_{\lambda, 24 \mu \rm m}$ for the 157 AGNs in Sample A that have MIPS \mbox{24 $\mu$m} detections. There is no apparent correlation between $L_{\rm X}$ and $\lambda L_{\lambda, 24 \mu \rm m}$ for these AGNs, which is consistent with the X-ray and \mbox{24 $\mu \rm m$} emission not being produced by the same process, suggesting that, typically, these AGNs do not significantly affect the observed \mbox{24 $\mu \rm m$} fluxes that are used to compute the SFRs.

\subsubsection{Sample Overview}\label{overview}

\begin{figure}[ht]
\includegraphics[width=3.3in]{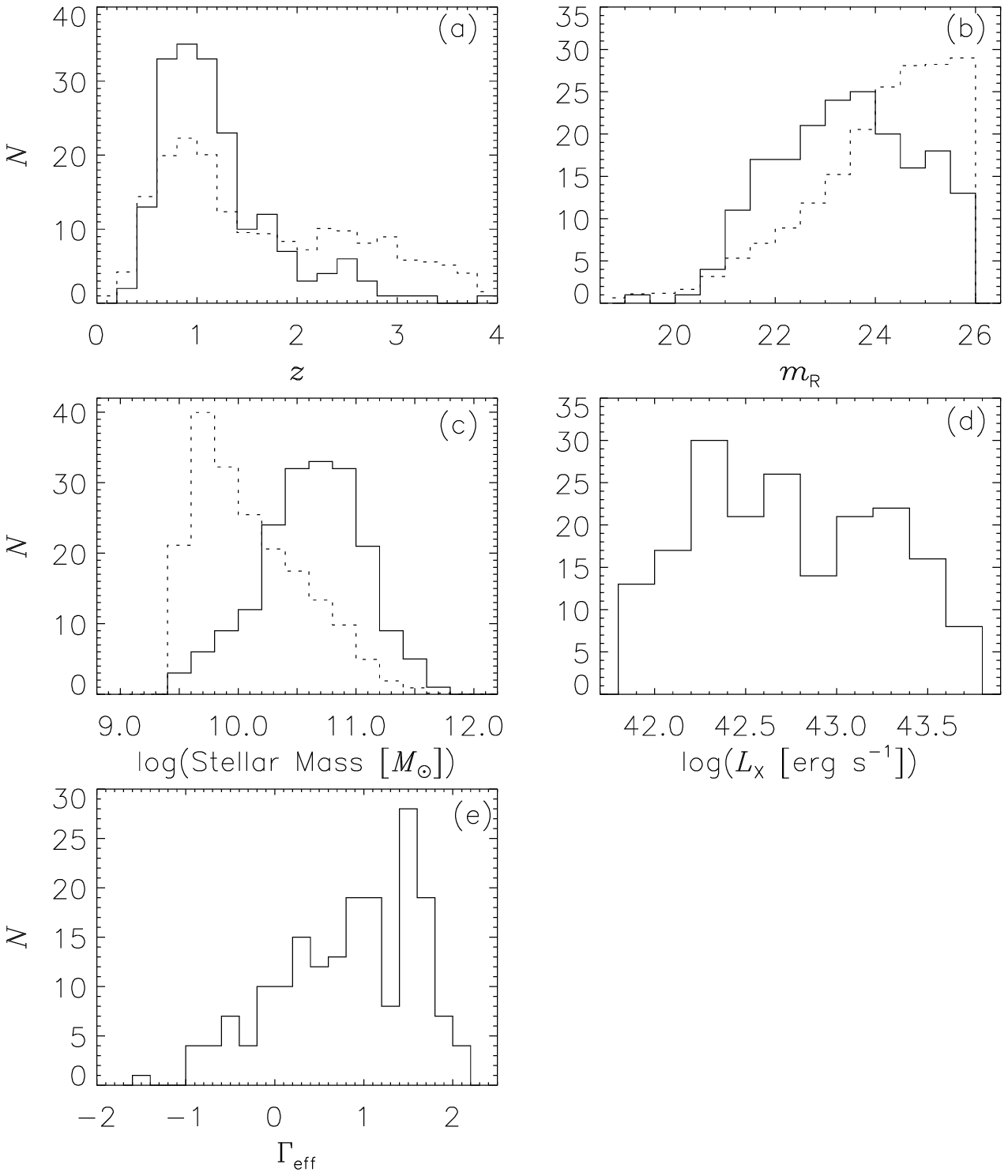}
\caption{Histograms of physical properties for \mbox{Sample A}, with AGNs shown as solid lines and non-AGN galaxies as dotted lines: (a) Redshift; (b) $R$-band magnitude;
(c) Stellar mass; (d) \mbox{0.5--8}~keV \mbox{X-ray} luminosity; and (e) Effective photon index.
The number of non-AGN galaxies has been rescaled to match that of AGNs in each plot.
The histogram of SFR is not shown, due to the existence of upper limits on SFR (see \S~\ref{sec_sfr}).
}
\label{histo}
\end{figure}

Table~\ref{all-data} presents relevant data
and Figure~\ref{histo} shows histograms of several physical properties for the sources in \mbox{Sample A}.
In \mbox{Sample A}, $\approx 95$\% ($\approx 98$\%) of the sources have rest-frame $K$-band ($I$-band) coverage or beyond,
which ensures the reliability of our stellar-mass estimates (see \S~\ref{stellarmass}). 
As seen in Fig.~\ref{histo}(c), it is apparent that AGN hosts are generally more massive than non-AGN galaxies. 

\begin{table*}[ht]
\caption{Relevant Data for the Sources in Sample A}
\resizebox{\textwidth}{!}{%
\begin{tabular}{ccccccccccccc}\hline\hline
No. & RA$_{\rm J2000.0}$ & DEC$_{\rm J2000.0}$ & $z$ & $m_{\rm R}$ & $M_{\rm U}$ & $M_{\rm B}$ & $M_{\rm V}$ & log($M_{\rm B/V/R/I/J/H/K,\star}/M_\odot$) & F$_{\rm mass}$ & SFR ($M_\odot$ ${\rm yr}^{-1}$) & XID & log[$L_{\rm X}/({\rm erg\:s^{-1}})$]  \\
(1) & (2) & (3) & (4) & (5) & (6) & (7) & (8) & (9) & (10) & (11) & (12) & (13)\\\hline
1 & 189.37823 & 62.18428 &  2.486 & 25.33 & $-$21.73 & $-$21.68 & $-$22.01 & (10.12, 10.13, 10.11, 10.08,  9.95,  9.89,  9.91) & 7 & 8.7 & 0 & -42.85\\
2 & 189.37852 & 62.18563 &  0.681 & 22.13 & $-$20.30 & $-$20.09 & $-$20.53 &  (9.74,  9.75,  9.76,  9.80,  9.71,  9.69,  9.63) & 7 & $-$1.8 & 0 & -41.46\\
3 & 189.38080 & 62.18413 &  1.169 & 24.49 & $-$20.38 & $-$20.39 & $-$20.80 &  (9.78,  9.79,  9.77,  9.74,  9.63,  9.59,  9.61) & 7 & 6.0 & 0 & -42.02\\
4 & 189.26192 & 62.18623 & +0.559 & 22.20 & $-$19.77 & $-$20.01 & $-$20.80 & (10.49, 10.50, 10.53, 10.56, 10.46, 10.43, 10.39) & 7 & 6.0 & 0 & -41.03\\
5 & 189.26080 & 62.18484 & +1.014 & 23.72 & $-$20.50 & $-$19.94 & $-$20.42 & (9.76, 9.76, 9.79, 9.85, 9.77, 9.75, 9.70) & 7 &    3.5 & 0 & -41.67\\
\hline
\end{tabular}}
\label{all-data}
\\This table is available in its entirety in a machine-readable form in the online journal. A portion is shown here for guidance regarding its form and content. The full table contains 4941 entries.
Columns:
(1) Source sequence number.
(2, 3) J2000 right ascension and declination of the optical counterpart (in degrees), respectively.
(4) Spectroscopic redshift (with a plus sign) or photometric redshift (without a plus sign).
(5) Apparent $R$-band magnitude (AB mags).
(6, 7, 8) Rest-frame absolute Johnson $U$-, $B$-, and $V$-band magnitude (Vega mags), respectively.
(9) Stellar-mass estimate derived at the $B$-, $V$-, $R$-, $I$-, $J$-, $H$-, and $K$-band (irregardless of whether there is corresponding rest-frame coverage of a source), respectively. For the 4941 sources in \mbox{Sample A},
the median of $(M_{\rm B/V/R/I/J/H,\star})/M_{\rm K,\star}$ is $\approx$ (1.46, 1.49, 1.48, 1.51, 1.19, 1.08), respectively.
(10) Flag of reddest rest-frame coverage of a source. This flag shows which stellar-mass estimate is adopted as the final estimate for a source (see \S~\ref{stellarmass}). F$_{\rm mass}=$(1, 2, 3, 4, 5, 6, 7) means that $M_{\rm B/V/R/I/J/H/K,\star}$ is adopted for a source, respectively.
(11) Star-formation rate (positive value) or upper limit on star-formation rate (negative value).
(12) X-ray ID from Alexander et al. (2003) for the CDF-N AGNs (XID$>0$) and Luo et al. (2008) for the CDF-S AGNs (XID$<0$), respectively (XID=0 for non-AGN galaxies).
Four CDF-N AGNs have XID$>503$ and are from the CDF-N supplementary optically bright {\it Chandra} source catalog; their corresponding XIDs in this supplementary catalog are XID$-$503.
Two CDF-S AGNs have $462<$XID$\leq 548$ (XID$>548$) and are from the supplementary CDF-S plus E-CDF-S {\it Chandra} source catalog (the CDF-S supplementary optically bright {\it Chandra} source catalog); their corresponding XIDs in this supplementary catalog are XID$-$462 (XID$-$548).
(13) Intrinsic \mbox{0.5--8}~keV X-ray luminosity (positive value; for AGNs) or upper limit on intrinsic \mbox{0.5--8}~keV X-ray luminosity (negative value; for non-AGN galaxies).
\end{table*}

In \mbox{Sample A}, $\approx 23$\% ($\approx 16$\%) of the galaxies (AGNs) have no MIPS 24 $\mu$m detection and were given upper limits on SFR.
At $0<z\leq 1$ none of the sources without MIPS 24 $\mu$m detections has an upper limit on SFR greater than \mbox{10 $M_\odot$ yr$^{-1}$}, and
at $1<z\leq 2$ ($2<z\leq 3$) only two of the galaxies without MIPS 24 $\mu$m detections have upper limits on SFR greater than 20 (30) $M_\odot$ yr$^{-1}$;
these indicate that \mbox{Sample A} is roughly complete above SFR$\approx 10$ (20, 30) \mbox{$M_\odot$ yr$^{-1}$} at $0<z\leq 1$ ($1<z\leq 2$, $2<z\leq 3$).

Figure~\ref{lum-z}(a) shows \mbox{0.5--8}~keV luminosities of AGNs in \mbox{Sample A} as a function of redshift. 
According to Fig.~\ref{lum-z}(a), AGNs in \mbox{Sample A} are complete up to $z\approx 1.5$, whereas the completeness drops toward higher redshifts;
above ${\rm log}(L_{\rm X})\approx 43.0$, AGNs in \mbox{Sample A} are roughly complete out to $z\approx 3$.
We note that the incompleteness of \mbox{X-ray} AGNs at high redshifts
does not affect AGN fraction calculations using the method detailed in \S~\ref{cmd1}.

\section{Results}\label{results}

We chose to use the rest-frame $U-V$ color (i.e., \mbox{$M_{\rm U}-M_{\rm V}$}) throughout this section
because this color straddles the \mbox{4000 $\AA$} break of the continuum and better tracks the age and metallicity variations of
the stellar populations in galaxies than the $U-B$ (i.e., $M_{\rm U}-M_{\rm B}$) color 
(e.g., Bell et al. 2004). We have confirmed that the same basic results can be obtained 
by using the rest-frame $U-B$ color.

\subsection{Color Bimodality}\label{sec:bi}

In this sub-section, we examine whether there is color bimodality (i.e., separation of the red sequence and the blue cloud) for non-AGN galaxies and AGN hosts in \mbox{Sample A}.

\subsubsection{Color Bimodality: Non-AGN Galaxies}\label{sec:bi-gal}

\begin{figure*}[ht]
\includegraphics[width=7.1in]{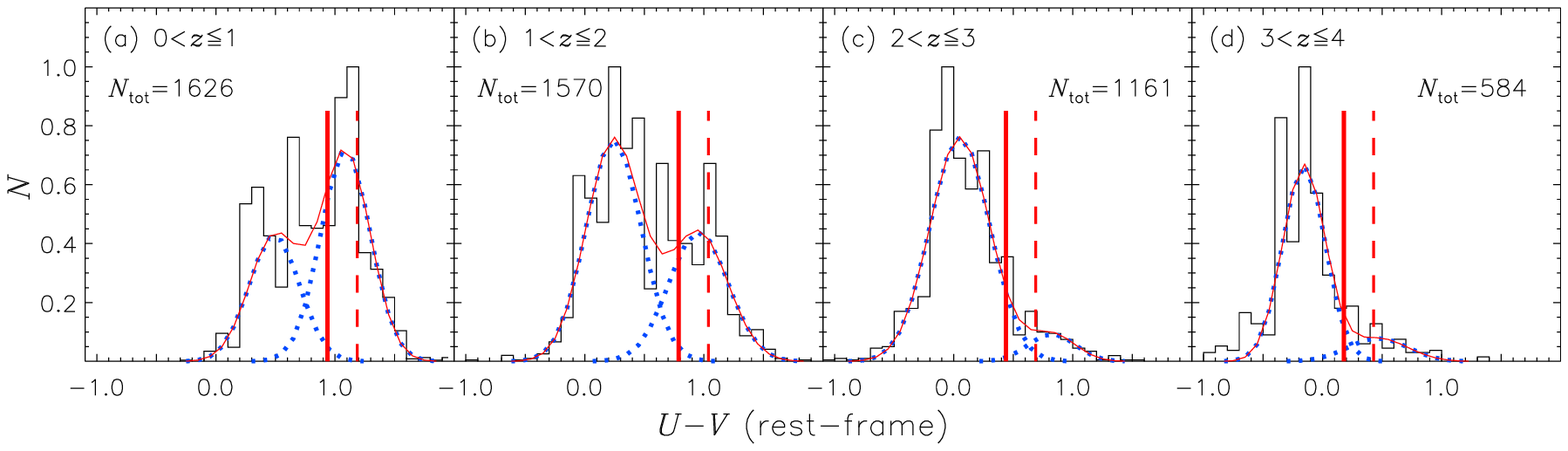}
\caption{Histograms of rest-frame $U-V$ color for non-AGN galaxies in \mbox{Sample A} (shown as black solid lines) in different redshift intervals,
with a bin size of $\Delta (U-V)=0.1$.
The peak value of each histogram ($N_{\rm peak}$) has been rescaled to unity
and the total number ($N_{\rm tot}$) of non-AGN galaxies in each subsample is shown.
Thick blue dotted lines show the two best-fit Gaussian components,
with their sum shown as the red solid line.
Best-fit Gaussian parameters and relevant statistical properties are shown in Table~\ref{tab:bi}.
A set of separation lines between the red sequence (to the right) and the blue cloud (to the left) is shown as vertical solid lines that are derived using Eqn.~(\ref{separation-line}) (see \S~\ref{sec:bi-gal}).
As a comparison, a set of lines representing the color-magnitude relation of red-sequence galaxies is shown as vertical dashed lines,
which are roughly 0.25 mag redward of the separation lines (Bell et al. 2004). (A color version of this figure is available in the online journal.)
}
\label{bi}
\end{figure*}

Figure~\ref{bi} shows histograms of rest-frame $U-V$ color for non-AGN galaxies in \mbox{Sample A}, 
which spans a broad range of about $-1.0\lsim U-V\lsim 2.0$.
The histograms were made with a set of redshift bins of \mbox{$\Delta z=1.0$} over \mbox{$z=0$--4}.
The choice of \mbox{$\Delta z=1.0$} facilitates examination of possible color evolution over cosmic time, avoids likely dilution of potential bimodal color behaviors when a wide range of redshift is considered (e.g., \mbox{$\Delta z=2.0$}), and avoids poor statistics due to low source counts in a narrow redshift bin (e.g., \mbox{$\Delta z=0.2$}).

We find that, in all four redshift ranges, the histogram of $U-V$ color for non-AGN galaxies can be acceptably fitted (using $\chi^2$ fitting) by two Gaussian components each in the form of 
$N=G_{\rm N}\hspace{0.5mm}{\rm exp}(-[(X-G_{\rm C})/G_{\rm W}]^2/2)$, where $X=U-V$, $G_{\rm N}$ is a normalization factor, and $G_{\rm C}$ ($G_{\rm W}$) is
the centroid (width) of the Gaussian component.\footnote{Based on
previous works that established galaxy color bimodality,
we applied an additional constraint of $0.1\leq G_{\rm W}\leq 0.5$ to the Gaussian fits, i.e., we required that
the width of the Gaussian component not be too small or too large.
We did not apply this additional constraint when examining possible color bimodality for AGN hosts because there is no established color bimodality for AGN hosts.}
The best-fit Gaussian parameters can be found in Table~\ref{tab:bi}. 
We note that a bimodal signature of the galaxy color distribution seems to be present in all four redshift ranges, although only a weak redward tail can be seen
in the highest range (see Fig.~\ref{bi}). 
We then tested for bimodality of the galaxy color distribution using the KMM mixture-modeling algorithm (e.g., Ashman, Bird, \& Zepf 1994).
This algorithm indicates bimodal color distributions with high statistical significance in all four redshift ranges shown in Fig.~\ref{bi}, 
with the derived KMM $P$-values all being equal to zero (see Table~\ref{tab:bi} for the $P$-values and corresponding mean values of two potential sub-populations).
However, we caution that the above seemingly established color bimodality is based on an assumption of Gaussian-distributed populations, 
which is unlikely to be the true distribution.

\begin{table*}[ht]
\caption{Best-Fit Gaussian Parameters and Relevant Statistical Properties for Figs.~\ref{bi}, \ref{bi2}, \& \ref{bi-av}}
\resizebox{\textwidth}{!}{%
\begin{tabular}{lcrrrrcccc}\hline\hline
       &     &            &             & Gaussian 1 & Gaussian 2 & KMM-test & $|G_{\rm C1}-G_{\rm C2}|/$ & & $F$-test \\
Figure & $z$ & $N_{\rm tot}$ & $N_{\rm peak}$ & ($G_{\rm C1}$, $G_{\rm W1}$, $G_{\rm N1}$) & ($G_{\rm C2}$, $G_{\rm W2}$, $G_{\rm N2}$) & ($P$-value, mean 1, mean 2) & $(G_{\rm W1}^2+G_{\rm W2}^2)^{0.5}$ & $G_{\rm N1}/G_{\rm N2}$ & $p$ \\
(1) & (2) & (3) & (4) & (5) & (6) & (7) & (8) & (9) & (10) \\\hline
Fig.~\ref{bi}(a) & $0.0<z\leq 1.0$ & 1626 & 230 & (0.490, 0.213, 0.426) & (1.088, 0.209, 0.716) & (0.000, 0.478, 1.112) & 2.01 & 0.60 & 0.0381\\
Fig.~\ref{bi}(b) & $1.0<z\leq 2.0$ & 1570 & 195 & (0.242, 0.231, 0.753) & (0.955, 0.248, 0.439) & (0.000, 0.272, 0.981) & 2.10 & 1.72 & 0.0029\\
Fig.~\ref{bi}(c) & $2.0<z\leq 3.0$ & 1161 & 200 & (0.052, 0.251, 0.762) & (0.824, 0.196, 0.094) & (0.000, 0.045, 0.799) & 2.42 & 8.14 & 0.0256\\
Fig.~\ref{bi}(d) & $3.0<z\leq 4.0$ &  584 & 133 &($-$0.156, 0.178, 0.667) & (0.465, 0.250, 0.079) & (0.000, $-$0.189, 0.600) & 2.02 & 8.45 & 0.2407\\ \hline
Fig.~\ref{bi2}(a) & $0.0<z\leq 1.0$ & 83    &   17 &(1.014, 0.481, 0.318) & (1.091, 0.056, 0.899) & (0.003, 0.489, 1.081)& 0.16 & 0.35 & 0.0513\\
Fig.~\ref{bi2}(b)& $1.0<z\leq 2.0$ & 85   &   11 & (0.537, 0.554, 0.259) & (0.982, 0.205, 0.681) & (0.002, 0.130, 0.940)& 0.75 & 0.38 & 0.2866 \\ \hline
Fig.~\ref{bi-av}(a) & $0.0<z\leq 1.0$ & 1626 & 197 & (0.088, 0.167, 0.561) & (0.759, 0.273, 0.826) & (0.000, 0.153, 0.804) & 2.09 & 0.68 & 0.0000\\
Fig.~\ref{bi-av}(b) & $1.0<z\leq 2.0$ & 1570 & 179 & (0.066, 0.204, 1.001) & (0.727, 0.266, 0.510) & (0.000, 0.099, 0.778) & 1.98 & 1.96 & 0.0000\\
Fig.~\ref{bi-av}(c) & $2.0<z\leq 3.0$ & 1161 & 232 & ($-$0.042, 0.225, 0.733) & (0.648, 0.214, 0.096) & (0.000, $-$0.042, 0.699) & 2.22 & 7.61 & 0.0061\\
Fig.~\ref{bi-av}(d) & $3.0<z\leq 4.0$ &  584 & 125 &($-$0.244, 0.222, 0.695) & (0.466, 0.250, 0.039) & (0.000, $-$0.252, 0.600) & 2.13 & 17.85 & 0.3220\\
Fig.~\ref{bi-av}(A) & $0.0<z\leq 1.0$ & 83    &   14 &(0.069, 0.250, 0.152) & (0.790, 0.237, 0.852) & (0.002, 0.056, 0.789)& 2.09 & 0.178 & 0.0385\\
Fig.~\ref{bi-av}(B)& $1.0<z\leq 2.0$ & 85    &  16 & (0.450, 0.500, 0.360) & (0.751, 0.052, 0.700) & (0.003, 0.470, 0.553)& 0.60 & 0.51 & 0.0117 \\ \hline
\end{tabular}}
\label{tab:bi}
\\Columns:
(1) Subplot considered.
(2) Redshift range considered.
(3) Total source number in each subplot.
(4) Peak value of the histogram in each subplot.
(5, 6) Gaussian component (centroid, width, normalization) that corresponds to the blue and red population, respectively.
(7) Result of the KMM mixture-modeling algorithm that tests for bimodality.
A $P$-value of $\lsim 0.05$ is required for a bimodality claim.
Mean 1 and mean 2 are the mean values of two potential sub-populations.
(8) Quantity that measures the separation between the centroids of two Gaussian components.
A value of $\leq 0.707$ is required for a bimodality claim.
(9) Quantity that measures the relative strength between two Gaussian components.
A value ranging from 0.1 to 10 is required for a bimodality claim.
(10) Probability from the $F$-statistic that measures whether the requirement of a two-Gaussian-component fit rather than a single-Gaussian-component fit is statistically significant.
A value of $\lsim 0.05$ is required for a bimodality claim.
\end{table*}

To assess further the validity of the claim of color bimodality,
we imposed three additional necessary conditions (i.e., these three conditions must be satisfied in order for a bimodality claim):
(1) The quantity $|G_{\rm C1}-G_{\rm C2}|/\sqrt{G_{\rm W1}^2+G_{\rm W2}^2}$ should be equal or greater than 0.707 (i.e., $\sqrt{2}/2$),
where the value of 0.707 is obtained when $|G_{\rm C1}-G_{\rm C2}|=G_{\rm W1}=G_{\rm W2}$.
This quantity is a measure of separation between the centroids of two components.
(2) The quantity $G_{\rm N1}/G_{\rm N2}$ should be in a range roughly from 0.1 to 10, i.e., one component should not be negligible compared to the other.
(3) The requirement of a two-Gaussian-component fit (rather than a single-Gaussian-component fit) is statistically significant.
To address Condition (3), we calculated the $F$-statistic and its associated probability given ``new'' and ``old'' values of $\chi^2$ and degrees of freedom 
(i.e., values derived from two-Gaussian-component fits vs. those derived from single-Gaussian-component fits).
A low probability from the $F$-test indicates that two-Gaussian-component fits are better than single-Gaussian-component fits
at a high confidence level.
As shown in Table~\ref{tab:bi},
the cases of $0<z\leq 1$, $1<z\leq 2$, and $2<z\leq 3$ that have apparent bimodal signatures (see Fig.~\ref{bi}) satisfy the above three conditions,
while the case of $3<z\leq 4$ that has a less prominent bimodal signature (also see Fig.~\ref{bi}) fails the third condition
(i.e., the requirement of a two-Gaussian-component fit is not statistically significant).

As a final check, we calculated the separation lines between the red sequence and the blue cloud for different redshift ranges
using the following equation and plotted them as vertical solid lines in Fig.~\ref{bi}:
\begin{equation}
(U-V)_{\rm rest}=-0.31z-0.08M_{\rm V}-0.51.
\label{separation-line}
\end{equation}
This equation was derived by Bell et al. (2004) who studied the color distribution of $\approx$25,000 $R\lsim 24$ galaxies with \mbox{$0.2<z\leq 1.1$}.
These four separation lines shown in Fig.~\ref{bi} (median values of $z$ and $M_{\rm V}$ used for each range of redshift)
seem to describe our observational data reasonably well,  
taking into account a typical color scatter of $\lsim 0.2$ mag for the red sequence color-magnitude relation 
(see \S~4 in Bell et al. 2004).
The general agreement between the separation lines defined by Eqn.~(\ref{separation-line}) and our data suggests:
(1) the term of $-0.08M_{\rm V}$ in Eqn.~(\ref{separation-line}) may be valid up to at least $z\approx 3$ since 
the range of $M_{\rm V}$ of the galaxies in \mbox{Sample A} is similar to that of the Bell et al. galaxies;
(2) the term of $-0.31z$ may also be valid up to at least $z\approx 3$, which indicates that the definitions of the red sequence and the blue cloud
may evolve with redshift, i.e., galaxies may tend to be bluer as they are younger (i.e., at higher redshifts; also see \S~\ref{color-dep}).
We note that a careful extension of Eqn.~(\ref{separation-line}) out to $z\approx 4$ would be useful, but is beyond the scope of this work. 

Based on the above analyses, we conclude that there is evidence that galaxy color bimodality holds from the local universe up to \mbox{$z\approx 3$},
with likely hints of color bimodality up to \mbox{$z\approx 4$}.
We note that the relative fraction of galaxies in the red and blue populations evolves strongly, i.e., the fraction of red-sequence galaxies generally decreases as the redshift increases 
(we caution that this trend is affected quantitatively by the incompleteness of red galaxies in \mbox{Sample A} at high redshifts);
and that the colors of both the blue and red components become redder from the early universe to the present time (see Fig.~\ref{bi} and the values of $G_{\rm N1}/G_{\rm N2}$, $G_{\rm C1}$, and $G_{\rm C2}$ in Table~\ref{tab:bi}).
Our result confirms many previous works on galaxy color bimodality 
(e.g., Strateva et al. 2001, Hogg et al. 2002, and Baldry et al. 2004 for the local universe;
Bell et al. 2004, Weiner et al. 2005, Willmer et al. 2006, Cirasuolo et al. 2007, Franzetti et al. 2007, and Taylor et al. 2009a for up to \mbox{$z\approx 1$--2};
Kriek et al. 2008 for the detection of a red sequence of massive field galaxies at $z\approx 2.3$; Brammer et al. 2009 for up to $z\approx 2.5$;
Giallongo et al. 2005 for up to \mbox{$z\approx 2.5$--3})
and extends them to higher redshifts.
However, we note that some authors found no color bimodality for galaxies at high redshifts 
(e.g., Cirasuolo et al. 2007 for $z\gsim 1.5$; Labb\'{e} et al. 2007 for no well-defined red sequence at $z\approx 3$). 
Therefore, the situation is controversial at $z\gsim 1.5$, unlike the well-established galaxy color-bimodality below $z\approx 1.5$.

\subsubsection{Color Bimodality: AGN Hosts}\label{sec:bi-agn}

\begin{figure*}[ht]
\includegraphics[width=7.1in]{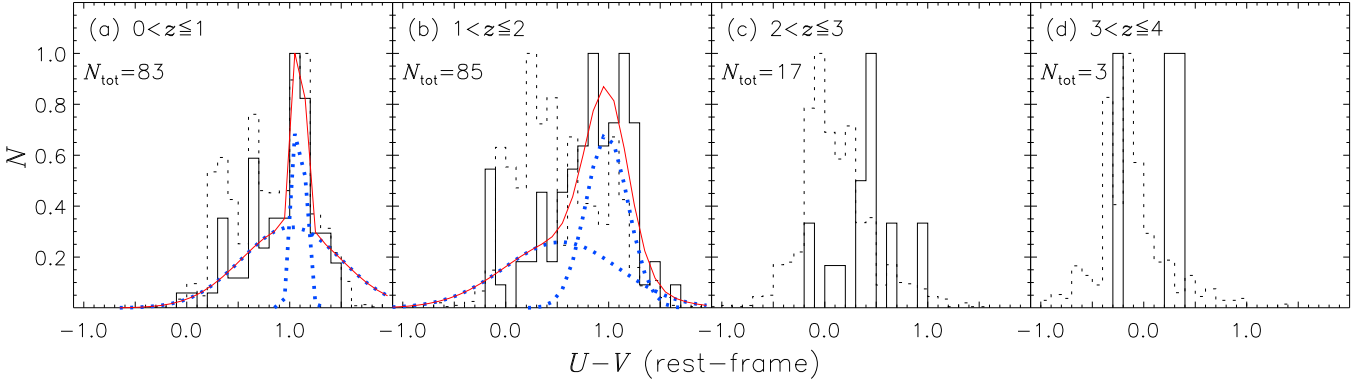}
\caption{Same as Fig.~\ref{bi}, but for AGN hosts (shown as black solid lines) in \mbox{Sample A}.
For comparison, histograms for non-AGN galaxies (from Fig.~\ref{bi}) are also shown as the dotted lines.
Note that Gaussian fits were applied only to Panels (a, b) which have a sufficient number of AGNs for meaningful analysis.
Best-fit Gaussian parameters and relevant statistical properties for these two panels are also shown in Table~\ref{tab:bi}.
AGN hosts are generally redder than non-AGN galaxies. (A color version of this figure is available in the online journal.)}
\label{bi2}
\end{figure*}

We performed the same procedure as for the case of non-AGN galaxies to test for color bimodality of AGN hosts in \mbox{Sample A}.  
The results are shown in Fig.~\ref{bi2} and Table~\ref{tab:bi}.
We note that only two redshift ranges were considered ($0.0<z\leq 1.0$ and $1.0<z\leq 2.0$);
the redshift ranges of $2.0<z\leq 3.0$ and $3.0<z\leq 4.0$ were not considered due to an insufficient number of sources for meaningful analysis.
For $0.0<z\leq 1.0$, the condition of $|G_{\rm C1}-G_{\rm C2}|/\sqrt{G_{\rm W1}^2+G_{\rm W2}^2}\geq 0.707$ is not satisfied (a value of 0.16 was obtained)
and the requirement of a two-Gaussian-component fit rather than
a single-Gaussian-component fit is not statistically significant
($p=0.0513$ in the \mbox{$F$-test}).
For $1.0<z\leq 2.0$, the requirement of a two-Gaussian-component fit is not statistically significant ($p=0.2866$ in the \mbox{$F$-test}).
Therefore, our result reveals no apparent color bimodality for AGN hosts, which is in agreement with previous studies 
(e.g., B\"{o}hm \& Wisotzki 2007 for a sample with a mean redshift of $\langle z \rangle\approx 0.6$; Silverman et al. 2008b for \mbox{$0.4\leq z\leq 1.1$})
and extends them to \mbox{$z\approx 1$--2}. 

\begin{figure*}[ht]
\includegraphics[width=7.1in]{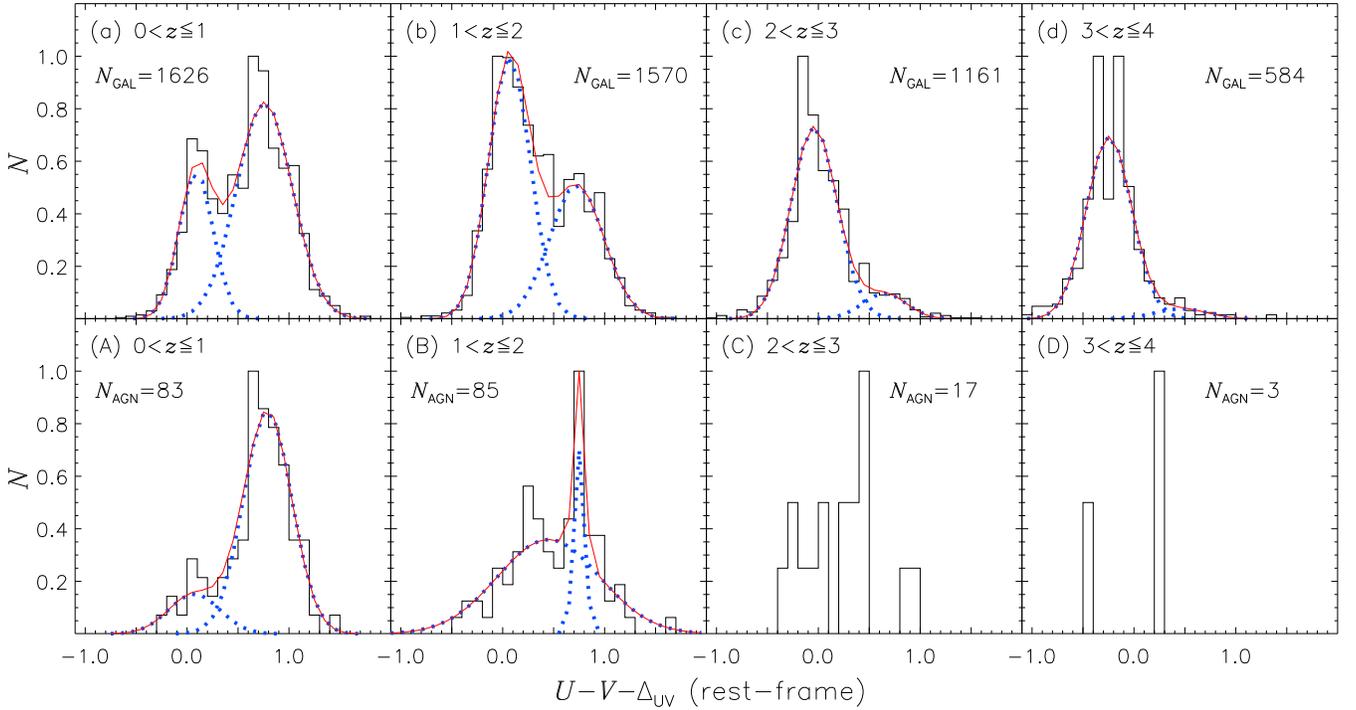}
\caption{Same as Figs.~\ref{bi} and \ref{bi2}, but correcting for dust extinction for non-AGN galaxies (Panels (a -- d)) and AGN hosts (Panels (A -- D)) in \mbox{Sample A}, respectively.
Best-fit Gaussian parameters and relevant statistical properties are also shown in Table~\ref{tab:bi}. (A color version of this figure is available in the online journal.)}
\label{bi-av}
\end{figure*}

According to Fig.~\ref{bi2}, it also seems clear that AGN hosts are generally redder than non-AGN galaxies up to $z\approx 3$--4.
This result confirms previous works (e.g., Barger et al. 2003; Nandra et al. 2007; Brusa et al. 2009; Treister et al. 2009; but see \S~\ref{color-dep} and \ref{mass-effect}).

\subsubsection{Color Bimodality: Effects of Dust Reddening}\label{dust}

Recent studies have investigated the effects of dust reddening when analyzing galaxy colors (e.g., Cowie \& Barger 2008; Brammer et al. 2009) and find that
many of the green-valley sources in the CMD are dust-reddened blue-cloud sources, rather than sources undergoing a blue-to-red transition.
Specifically, Brammer et al. (2009) studied a $K$-band selected ($K<22.8$) galaxy sample and argued that 
correcting the rest-frame colors for dust reddening allows an improved
separation between the red sequence and the blue cloud up to $z\approx 2.5$. 
To examine the effects of dust reddening on \mbox{Sample A},
we plot the dust-corrected color (i.e., $U-V-\Delta_{\rm UV}$) histograms for both non-AGN galaxies and AGN hosts in \mbox{Sample A} in Fig.~\ref{bi-av}.
For this assessment, we followed \mbox{Eqn. (2)} of Brammer et al. (2009), $\Delta_{\rm UV}=0.47 A_{\rm V}$, to correct for dust reddening.
We obtained $V$-band extinction information (i.e., $A_{\rm V}$) from the template SED fitting (see \S~\ref{absmag}). 

According to Fig.~\ref{bi-av} and Table~\ref{tab:bi},
it seems, after correcting for dust extinction, that: 
(1) For non-AGN galaxies at $0<z\leq 1$, $1<z\leq 2$, and $2<z\leq 3$, there is apparent improvement of the separation between the red sequence and the blue cloud,
in the sense that the profiles of the red and blue components appear smoother and clearer and the $p$ values from the \mbox{$F$-test} become significantly smaller; but
for non-AGN galaxies at $3<z\leq 4$, there is still no apparent color-bimodality (both the value of $G_{\rm N1}/G_{\rm N2}$ and the $p$ value from the \mbox{$F$-test} are too large).
(2) For AGN hosts at $0<z\leq 1$, there appears to be some improvement of the separation between the red sequence and the blue cloud.
Following the procedure described in \S~\ref{sec:bi-gal},
we find likely evidence of color bimodality for AGN hosts at $0<z\leq 1$ (we caution that the relatively weak blue component could simply be a blueward skewed tail of the red component).
For AGN hosts at $1<z\leq 2$, we still find no evidence of color bimodality (the condition of $|G_{\rm C1}-G_{\rm C2}|/\sqrt{G_{\rm W1}^2+G_{\rm W2}^2}\geq 0.707$ is not satisfied; a value of 0.60 was obtained).
(3) Generally, AGN hosts still appear redder than non-AGN galaxies at all redshift ranges (comparing Figs.~\ref{bi-av}(a -- d) with Figs.~\ref{bi-av}(A -- D), respectively).

We find that there are systematic blueward shifts in the colors due to de-reddening by
comparing Figs.~\ref{bi} and \ref{bi2} with Fig.~\ref{bi-av},
which suggests that dust extinction could be responsible for the red colors of dusty star-forming galaxies including many AGN hosts (also see, e.g., Brusa et al. 2009).
However, based on the above analysis, we conclude that dust extinction should not affect our results on color-bimodality in a significant way.
We have also verified that the results obtained hereafter will be unchanged qualitatively after using the above dust de-reddening technique.
Furthermore, most previous CMD-related works are based on dust-uncorrected colors, e.g., the separation line between the red sequence and the blue cloud (i.e., Eqn.~(\ref{separation-line})).
Therefore, we adopted dust-uncorrected colors hereafter for the purpose of straightforward comparison with previous works. 
We note that a more careful assessment of the effects of dust reddening is desirable in the future, but is beyond the scope of this work.

\subsection{Color-Magnitude Relations}\label{sec:cmd}

In this section,
we first investigate color-magnitude relations of AGN hosts and non-AGN galaxies;
we then explore the location of luminous submillimeter galaxies in the CMD and color-mass diagram. 

\subsubsection{CMDs for AGN Hosts and Non-AGN Galaxies}\label{cmd1} 

\begin{figure*}[ht]
\includegraphics[width=7.1in]{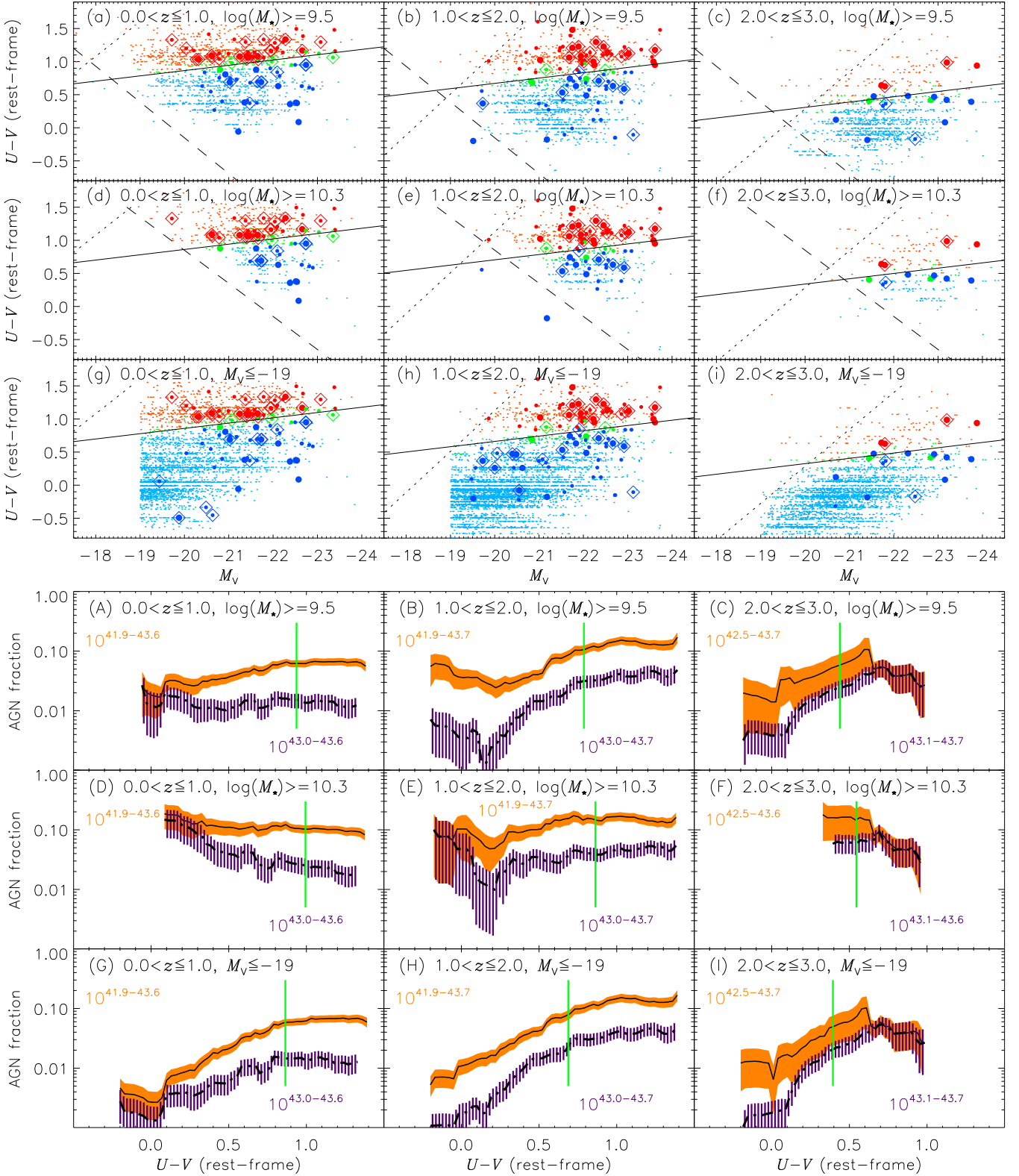}
\caption{(a -- i) Color-magnitude diagrams for Samples B, C, and D: rest-frame $U-V$ colors are shown against $V$-band absolute magnitudes, with AGN hosts shown as filled circles and non-AGN galaxies as small dots.
Large filled circles indicate \mbox{X-ray} luminous AGNs with $L_{\rm X} \geq 10^{43}$ \mbox{erg s$^{-1}$}. Diamonds indicate \mbox{X-ray} hard AGNs with $\Gamma_{\rm eff} \leq 0.5$.
Red/green/blue symbols represent sources in the regions of red sequence/green valley/blue cloud (see \S~\ref{sec:cmd}).
Note that the overlaps between the regions of red sequence/green valley/blue cloud are caused by our source-by-source classification scheme (see \S~\ref{sec:cmd}).
The solid lines show the color limits that separate the red sequence and the blue cloud, derived using Eqn.~(\ref{separation-line}) and the median redshift of galaxies for each subsample.
The dashed lines correspond roughly to the limit imposed by the stellar-mass cut of either $M_\star \geq 10^{9.5} M_\odot$ or $M_\star \geq 10^{10.3} M_\odot$.
The dotted lines roughly show the flux limit of $m_{\rm R}\leq 26$.
Note that the apparent horizontal stripes in the plots are due to the fact that
the galaxies on the same stripe have the same (or very similar) best-fit template but different redshifts,
which leads to the same (or very similar) $U-V$ colors but different values of $M_{\rm V}$.
(A -- I) AGN fraction as a function of rest-frame $U-V$ color (in bins of $\Delta (U-V)=0.4$) corresponding to each of the above cases of (a -- i), respectively.
Black solid (dashed-dot) curves show the fractions of AGNs (X-ray luminous AGNs; i.e., $43.0\leq {\rm log}(L_{\rm 0.5-8\:keV}/({\rm erg\:s^{-1}}))\leq 43.7$); orange (purple) filled patterns show the corresponding $1\sigma$ confidence ranges 
for AGNs (X-ray luminous AGNs); and orange (purple) numbers show the ranges of X-ray luminosity for AGNs (X-ray luminous AGNs) considered.
The vertical lines roughly show the separation lines derived using Eqn.~(\ref{separation-line}) and the median redshift and $M_{\rm V}$ of galaxies for each case. (A color version of this figure is available in the online journal.)
}
\label{cmd}
\end{figure*}

The CMDs for the AGNs and their parent galaxies in Samples B, C, and D are shown in Figs.~\ref{cmd}(a -- i).
In each CMD, we divided sources (both non-AGN galaxies and AGN hosts) into three groups 
(the red sequence, the green valley, and the blue cloud) using Eqn.~(\ref{separation-line}) on a source-by-source basis.
If a source satisfies $(U-V)_{\rm rest}+0.31z+0.08M_{\rm V}+0.51>0.05$, it is in the red sequence;
if a source satisfies $-0.05\leq (U-V)_{\rm rest}+0.31z+0.08M_{\rm V}+0.51\leq 0.05$, it is in the green valley; and
if a source satisfies $(U-V)_{\rm rest}+0.31z+0.08M_{\rm V}+0.51<-0.05$, it is in the blue cloud.
Figures~\ref{cmd}(A -- I) show corresponding plots of AGN fraction as a function of rest-frame $U-V$ color for Samples B, C, and D.
For a given \mbox{X-ray} luminosity range, we computed the AGN fraction ($f$) and its associated error ($\sigma_f$) according to the following two equations that were used by Silverman et al. (2008):
\begin{equation}
f=\sum_{i=1}^{N}\frac{1}{N_{\rm gal,i}},
\label{fraction}
\end{equation}
\begin{equation}
\sigma_f^2\approx\sum_{i=1}^{N}\frac{1}{{N_{\rm gal,i}^{2}}},
\label{fraction_error}
\end{equation}
where $f$ and $\sigma_f$ are a sum over the full sample of AGNs ($N$) with ${N_{\rm gal,i}}$ denoting the number of galaxies capable of hosting the $i$th detectable AGN with X-ray luminosity $L_{\rm X}^i$. 
This method not only takes into account the spatially varying sensitivity limits of the 2~Ms CDF observations using the sensitivity maps derived in \S~4 of Luo et al. (2008),
but also treats the incompleteness of \mbox{X-ray} AGNs at high redshifts [see Fig.~\ref{lum-z}(a)] effectively so that it does not affect AGN fractions.

We have the following observations up to $z\approx 2$--3 of the material in Fig.~\ref{cmd}.
(1) The majority of AGNs reside in massive galaxies (comparing AGN counts in Figs.~\ref{cmd}(a -- c) with those in Figs.~\ref{cmd}(d -- f);
also see Fig.~\ref{mass-z} and Fig.~\ref{histo}(c); see \S~\ref{color-dep} for further relevant details).
(2) For \mbox{Sample C} (i.e., $M_\star \geq 10^{10.3} M_\odot$),
AGN hosts do not appear more luminous than non-AGN galaxies; 
both AGN hosts and non-AGN galaxies are luminous and share similar $M_{\rm V}$ distributions. 
This is in contrast to the result from either \mbox{Sample B} (i.e., $M_\star \geq 10^{9.5} M_\odot$) or \mbox{Sample D} (i.e., $M_{\rm V}\leq -19$)
that AGN hosts are generally more luminous than non-AGN galaxies (i.e., having more negative values of $M_{\rm V}$);
the latter result was also found by past CMD works (e.g., Nandra et al. 2007; Silverman et al. 2008b).
(3) For \mbox{Sample C}, AGNs do not seem to reside predominantly in the red sequence, the top of the blue cloud, or the green valley in between; in fact, 
AGN hosts seem to spread throughout the CMDs (see, particularly, Figs.~\ref{cmd}(d, e)).
This is different from the result obtained with either \mbox{Sample B} or \mbox{Sample D} that there is apparent AGN clustering in the CMDs;  
the latter result was also obtained by previous CMD studies (e.g., Nandra et al. 2007; Rovilos \& Georgantopoulos 2007;
Westoby et al. 2007; Silverman et al. 2008b; Hickox et al. 2009; Schawinski et al. 2010).
(4) For \mbox{Sample C}, the AGN fraction seems to remain broadly constant (at \mbox{$\approx 10$\%}) regardless of color [see Figs.~\ref{cmd}(D -- F)].
This is in contrary to the case of either \mbox{Sample B} or \mbox{Sample D}
where the AGN fraction generally increases as colors become redder, with likely signs of leveling off at red colors [see Figs.~\ref{cmd}(\mbox{A -- C}, \mbox{G -- I})]. 
Whether a large population of low-mass blue galaxies is included in the analysis holds the key to explaining the above different trends of AGN fraction seen between \mbox{Sample C} and \mbox{Sample B/D}, i.e.,
this population of galaxies is included in \mbox{Sample B/D},
which is responsible for the drop in AGN fractions at blue colors (also see Silverman et al. 2009).
Results on the fraction of X-ray luminous AGNs (i.e., $L_{\rm X}\geq 10^{43}$ \mbox{erg s$^{-1}$}) as a function of color (for Samples B, C, and D) are in general agreement with the above results on the AGN fraction.
We note that the results on AGN fraction for \mbox{Sample C} likely indicate the duty cycle of SMBH growth in typical massive galaxies.
(5) For Samples B, C, and D, it seems that the majority of \mbox{X-ray} hard AGNs (i.e., $\Gamma_{\rm eff} \leq 0.5$) reside in the red sequence, which we suspect is due to the fact that the majority of AGNs overall reside in the red sequence.
Indeed, Fisher's exact probability test shows no strong evidence [values of $P_{\rm Fisher}>0.1$ are obtained except for one case (Fig.~\ref{cmd}(a)) where $P_{\rm Fisher}=0.034$] that \mbox{X-ray} hard AGNs preferentially reside in the red sequence compared to AGNs overall.
We note that the incidence of \mbox{X-ray} hard AGNs in the red sequence was reported by previous works and has been used to argue for BH accretion persisting after the termination of star formation (e.g., Nandra et al. 2007; Georgakakis et al. 2008).

\subsubsection{CMDs for Luminous Submillimeter Galaxies}

Luminous submillimeter galaxies (SMGs) are a class of strongly star-forming, dust-obscured galaxies at \mbox{$z\approx 1$--3} 
that have been studied extensively
(e.g., Alexander et al. 2005a, 2005b; Chapman et al. 2005; Pope et al. 2006, 2008; Biggs et al. 2010; Laird et al. 2010; Swinbank et al. 2010; Wardlow et al. 2010)
but have yet to be put into the context of CMDs.
Since we can now extend CMD studies of AGN hosts and non-AGN galaxies into this redshift regime,
it is of interest to see where SMGs lie in CMDs,
which may provide one useful guide to interpreting CMD results.

The accuracy of submillimeter positions (for SMGs) is relatively poor (typically accurate to \mbox{$\approx 1$--7\arcsec}; see, e.g., Biggs et al. 2010), compared to those of optical, infrared, or X-ray positions. Thus it is challenging to identify secure counterparts for SMGs.
Chapman et al. (2005) and Biggs et al. (2010) have identified reliable radio and/or MIPS \mbox{24 $\mu$m} 
counterparts for samples of SMGs in the \mbox{CDF-N} and \mbox{E-CDF-S} (i.e., Extended-{\it Chandra} Deep Field-South; Lehmer et al. 2005), respectively, through careful analyses.  
We therefore cross-matched the optical positions of the sources in \mbox{Sample A} with the radio and/or MIPS \mbox{24 $\mu$m} positions of the \mbox{CDF-N} and \mbox{E-CDF-S} SMGs provided by Chapman et al. (2005) and Biggs et al. (2010), respectively (a maximum matching radius of 0.75\arcsec \hspace{0.8mm} is adopted; the associated false-match probability is \mbox{$\approx 3.9$\%}). 
We identified 11 SMGs with $1.0\leq z\leq 3.0$ (with a median redshift of $\approx 2$) in \mbox{Sample A}; 5 of these SMGs host AGNs.\footnote{For the \mbox{CDF-N}, Chapman et al. (2005) presented a spectroscopically identified sample of 22 SMGs, among which 19 are located within the central $r_{\rm encircled}=$8\arcmin \hspace{0.5mm} radius area;
of these 19 SMGs, 15 have counterparts in our North base catalog and only 11 are in \mbox{Sample A}. 
For the \mbox{E-CDF-S}, Biggs et al. (2010) presented a submillimeter-flux-limited sample of 126 SMGs, among which 12 are located within the central $r_{\rm encircled}=$8\arcmin \hspace{0.5mm} radius area;
of these 12 SMGs, 5 have counterparts in our South base catalog and only 2 are in \mbox{Sample A}.
We thus have a total of 13 SMGs in \mbox{Sample A}; 11 of them have $1.0\leq z\leq 3.0$.}

\begin{figure*}[ht]
\includegraphics[width=7.1in]{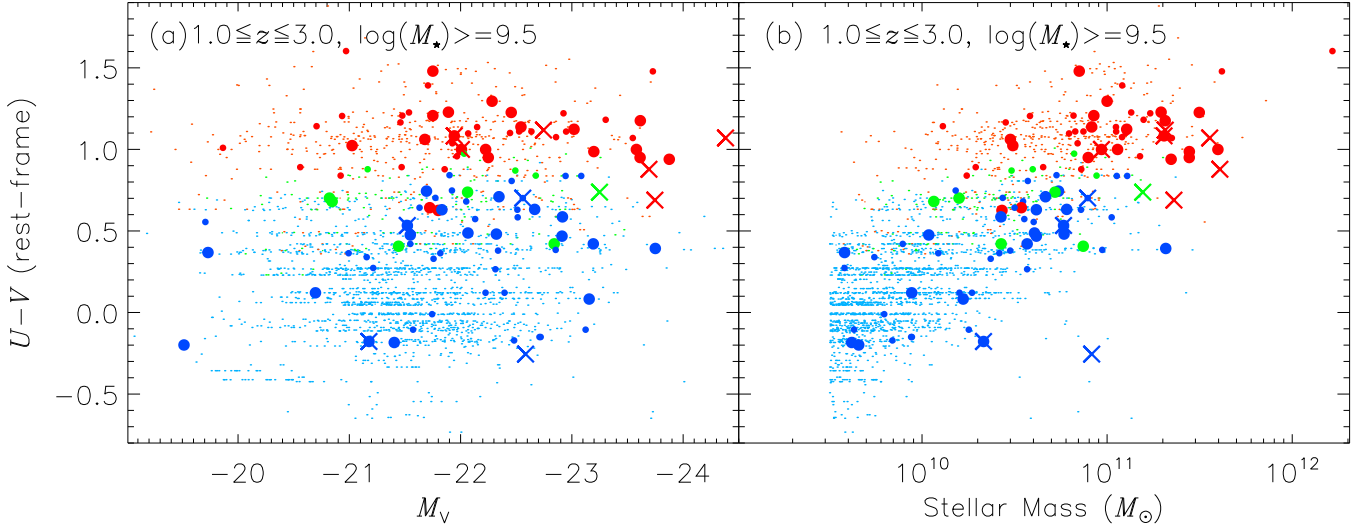}
\caption{Submillimeter galaxies (shown as crosses) in the color-magnitude diagram [Panel (a)] 
and color-mass diagram [Panel (b)] for \mbox{Sample A} at $1.0\leq z\leq 3.0$
(symbols and colors have the same meaning as those in Fig.~\ref{cmd}).
(A color version of this figure is available in the online journal.)}
\label{smg}
\end{figure*}

Figure~\ref{smg} shows the location of these 11 SMGs in the CMD and color-mass diagram.
These SMGs are spread throughout the CMD in the sense that they are found in the red sequence, the blue cloud, and the green valley in between, although on average they appear more luminous than the general galaxy population (SMGs with SFRs of $\approx 500$ $M_\odot$ yr$^{-1}$ would have \mbox{$L_{\rm UV}\approx 5\times 10^{12}L_\odot$} if they were not obscured).
This is likely due to extinction effects in these extreme systems:
the red colors might well indicate dust extinction rather than old stellar populations; and  
the blue colors might reflect that a small fraction of the star  
formation is unobscured and thus dominates over the obscured star-formation component 
in the observed optical--near-IR band.
Therefore, the observed colors of these SMGs do not represent their intrinsic colors.
In the color-mass diagram, these SMGs, as expected, lie toward the high stellar-mass end (i.e., they are much more massive than the general galaxy population);
the median stellar mass of these SMGs is $\approx 1.6\times 10^{11} M_\odot$, which is toward the low end of the estimates from Borys et al. (2005).
The location of SMGs in the CMD and color-mass diagram 
provides a powerful and specific example of how CMD and color-mass diagram
results can be subject to reddening effects,
which underscores the need for caution when interpreting the color-magnitude properties of small samples of extreme systems.

\subsection{Color Dependence on Physical Properties}\label{color-dep}

Measurements of the dependences of galaxy colors on physical properties such as redshift, SFR, and stellar mass can provide constraints on
models of galaxy formation and evolution.
In this sub-section, we examine the dependences of both AGN host colors and non-AGN galaxy colors on redshift, SFR, and stellar mass, 
and also examine the likely correlations between them.
%
The dependences of AGN host colors and non-AGN galaxy colors on redshift, SFR, and stellar mass for \mbox{Sample B} are shown in Fig.~\ref{cmd-z}
and the corresponding statistical properties are shown in Table~\ref{stat}.
Based on the information in Fig.~\ref{cmd-z} and Table~\ref{stat}, we obtain the following results:

\begin{figure*}[ht]
\includegraphics[width=7.1in]{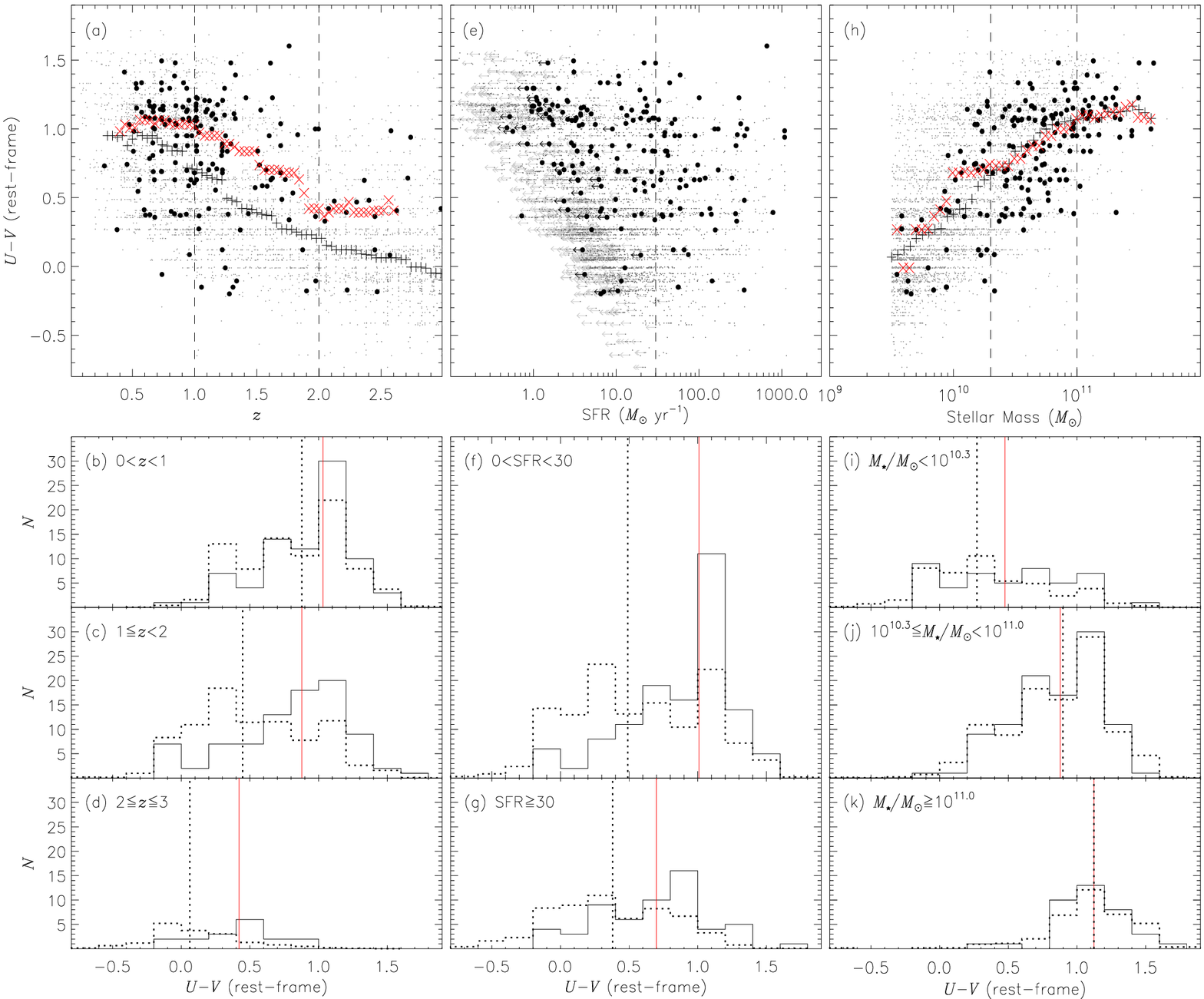}
\caption{Dependences of color on redshift (left panels), SFR (middle panels), and stellar mass (right panels) for the AGN hosts (filled circles) and non-AGN galaxies (small grey dots) in \mbox{Sample B}.
(a) Plot of rest-frame $U-V$ color vs. redshift. (b -- d) Histograms of rest-frame $U-V$ color for the AGN hosts (solid lines) and non-AGN galaxies (dotted lines) in three ranges of redshift.
(e) Plot of rest-frame $U-V$ color vs. SFR. The leftward arrows indicate upper limits on SFR.
(f, g) Histograms of rest-frame $U-V$ color for the AGN hosts (solid lines) and non-AGN galaxies (dotted lines) in two ranges of SFR. The $y$-axis in (f) has a range of 0--70.
(h) Color-mass diagram.
(i -- k) Histograms of rest-frame $U-V$ color for the AGNs (solid lines) and non-AGN galaxies (dotted lines) in three ranges of stellar mass.
In panels (a) and (h), the crosses (pluses) show the running medians of $U-V$ colors for AGN hosts (non-AGN galaxies) in bins of $\Delta z=0.4$ and $\Delta {\rm log}(M_\star)=0.5$, respectively. The running medians of $U-V$ colors were not computed in (e) due to the existence of upper limits on SFR.
In (a, e, h), the vertical dashed lines indicate the parameter values adopted to divide into subsamples.
For the histograms, non-AGN galaxy numbers have been rescaled to match those of AGN hosts; vertical solid (dotted) lines indicate median AGN host (galaxy) $U-V$ colors of each subsample. (A color version of this figure is available in the online journal.)}
\label{cmd-z}
\end{figure*}

\begin{table*}[ht]
\caption{Statistical Properties of Color Distributions for Fig.~\ref{cmd-z}}
\resizebox{\textwidth}{!}{%
\begin{tabular}{lrrrrrrrr}\hline\hline
Figure & \hspace{4.3mm}AGNs & \hspace{4.3mm}$(U-V)_{\rm AGN}$ & \hspace{4.3mm}$[r_{\rm S} (p_{\rm S})]_{\rm AGN}$ & \hspace{9.3mm}Galaxies & \hspace{4.3mm}$(U-V)_{\rm GAL}$ & \hspace{4.3mm}$[r_{\rm S} (p_{\rm S})]_{\rm GAL}$ & \hspace{4.3mm}K-S (\%) & \hspace{16mm}Condition \\
 & & & & & & & AGN-GAL \\
(1) & (2) & (3) & (4) & (5) & (6) & (7) & (8) & (9) \\\hline
\ref{cmd-z}(a) & 185 (100.0\%) & 0.89 & $-$0.33 (0.000) & 4357 (100.0\%) & 0.47 & $-$0.63 (0.000) & 0.0 & $0.0< z\leq 3.0$ \\
\ref{cmd-z}(b) & 83 (44.8\%) & 1.03 & --- & 1626 (37.3\%) & 0.88 & --- & 7.4 &  $0.0<z<1.0$ \\
\ref{cmd-z}(c) & 85 (46.0\%) & 0.88 & --- & 1570 (36.0\%) & 0.45 & --- & 0.0 & $1.0\leq z <2.0$ \\
\ref{cmd-z}(d) & 17 (9.1\%) & 0.42 & --- & 1161 (26.7\%) & 0.06 & --- & 0.1 & $2.0\leq z\leq 3.0$ \\\hline
\ref{cmd-z}(e) & 185 (100.0\%) & 0.89 & --- & 4357 (100.0\%) & 0.47 & --- & 0.0 & $0.0< z\leq 3.0$  \\
\ref{cmd-z}(f) & 127 (68.6\%) & 1.01 & --- & 3524 (80.9\%) & 0.49 & --- & 0.0 &  $0.0<{\rm SFR}<30.0$ \\
\ref{cmd-z}(g) & 58 (31.4\%) & 0.70 & --- & 833 (19.1\%) & 0.38 & --- & 0.1 &  ${\rm SFR}\geq 30.0$ \\\hline
\ref{cmd-z}(h) & 185 (100.0\%) & 0.89 & 0.57 (0.000) & 4357 (100.0\%) & 0.47 & 0.63 (0.000) & 0.0 & $0.0< z\leq 3.0$  \\
\ref{cmd-z}(i) & 46 (24.9\%) & 0.48 & --- & 2889 (66.3\%) & 0.27 & --- & 4.3 & $10^{9.5}\leq M_\star<10^{10.3}$ \\
\ref{cmd-z}(j) & 102 (55.1\%) & 0.88 & --- & 1269 (29.1\%) & 0.90 & --- & 26.7 &  $10^{10.3}\leq M_\star<10^{11.0}$ \\
\ref{cmd-z}(k) & 37 (20.0\%) & 1.12 & --- & 199 (4.6\%) & 1.12 & --- & 41.2 & $M_\star\geq 10^{11.0}$ \\\hline\hline
 & AGN-AGN & AGN-AGN & AGN-AGN & & GAL-GAL & GAL-GAL & GAL-GAL \\
 & (10) & (11) & (12) & & (13) & (14) & (15)\\ \hline
Figures & \ref{cmd-z}(b)-\ref{cmd-z}(c) & \ref{cmd-z}(b)-\ref{cmd-z}(d) & \ref{cmd-z}(c)-\ref{cmd-z}(d) & & \ref{cmd-z}(b)-\ref{cmd-z}(c) & \ref{cmd-z}(b)-\ref{cmd-z}(d) & \ref{cmd-z}(c)-\ref{cmd-z}(d) \\
$r_{\rm S} (p_{\rm S})$ & $-0.21$ (0.007) & $-0.34$ (0.001) & $-0.42$ (0.000) & & $-0.45$ (0.000) & $-0.67$ (0.000) & $-0.54$ (0.000) \\\hline
Figures & \ref{cmd-z}(i)-\ref{cmd-z}(j) & \ref{cmd-z}(i)-\ref{cmd-z}(k) & \ref{cmd-z}(j)-\ref{cmd-z}(k) & & \ref{cmd-z}(i)-\ref{cmd-z}(j) & \ref{cmd-z}(i)-\ref{cmd-z}(k) & \ref{cmd-z}(j)-\ref{cmd-z}(k) \\
$r_{\rm S} (p_{\rm S})$ & 0.49 (0.000) & 0.74 (0.000) & 0.46 (0.000) & & 0.60 (0.000) & 0.51 (0.000) & 0.32 (0.000) \\\hline\hline
\end{tabular}}
\\Columns:
(1) Subplot considered.
(2, 5) Number (percentage) of AGNs and non-AGN galaxies in each subplot, respectively.
(3, 6) Median $U-V$ color of AGNs and non-AGN galaxies in each subplot, respectively.
(4, 7) Spearman's rank correlation coefficient $r_{\rm S}$ and its associated probability $p_{\rm S}$ for AGNs and non-AGN galaxies, respectively (see Footnote~\ref{spearman}).
(8) Probability ($P_{\rm KS}$) that the color distributions of AGNs and non-AGN galaxies could be drawn from the same parent population, derived from a Kolmogorov-Smirnov (K-S) test (see Footnote~\ref{ks}).
(9) Relevant condition applied in each subplot.
(10 -- 12, 13 -- 15) The values of $r_{\rm S} (p_{\rm S})$ derived when combining AGNs or non-AGN galaxies in two subplots, respectively.
Note that we did not derive any $r_{\rm S} (p_{\rm S})$ for SFR-related cases since some sources have only upper limits on SFR.
\label{stat}
\end{table*}

(1) AGN hosts are generally redder than non-AGN galaxies irrespective of redshift or SFR range considered 
[see the running medians of colors plotted in Fig.~\ref{cmd-z}(a), the color distributions in Figs.~\ref{cmd-z}(b -- d, f, g),
and the $P_{\rm KS}$ values\footnote{$P_{\rm KS}$ is the probability that the two color distributions considered could be drawn from the same parent population, according to the null hypothesis from a Kolmogorov-Smirnov (K-S) test; a small value of $P_{\rm KS}$ (e.g., $\leq 5.0$\%) indicates that the two color distributions considered are different.\label{ks}} in Table~\ref{stat}].
However, we find that this trend seems to be valid only up to $M_\star\approx 10^{10.3}M_\odot$ when examining the colors as a function of stellar mass;
above $10^{10.3}M_\odot$, both AGN hosts and non-AGN galaxies have similar color distributions
[see the running medians of colors plotted in Fig.~\ref{cmd-z}(h), the color distributions in Figs.~\ref{cmd-z}(i -- k),
and the $P_{\rm KS}$ values in Table~\ref{stat}].
This latter observation appears to be inconsistent with the result from \S~\ref{sec:bi} that  
non-AGN galaxies are bimodal in color but AGN hosts are not (see Fig.~\ref{bi} and Fig.~\ref{bi2}).
We note that this apparent inconsistency is due to the fact that samples with different stellar-mass cuts (i.e., $M_\star\geq 10^{10.3}M_\odot$ vs. $M_\star\geq 10^{9.5}M_\odot$) were used respectively.
We thus made versions of Fig.~\ref{bi} and Fig.~\ref{bi2} for \mbox{Sample C} (i.e., with $M_\star\geq 10^{10.3}M_\odot$)
and find that the color distributions of non-AGN galaxies and AGN hosts are similar and neither shows apparent bimodality. 

(2) There is an apparent color evolution as a function of redshift such that the higher the redshift, the bluer the colors, for both AGN hosts and non-AGN galaxies 
(see \mbox{Figs.~\ref{cmd-z}(a -- d)} and the values of Spearman's rank correlation coefficient ($r_{\rm S}$) as well as the associated probability ($p_{\rm S}$)\footnote{$p_{\rm S}$ is the two-sided significance of deviation from zero and a small $p_{\rm S}$ value indicates that the result of $r_{\rm S}$ is obtained at a high significance level (i.e., is statistically meaningful). In this paper, we adopt the following criteria: \mbox{$|r_{\rm S}|\approx 0.0$--0.3} indicates no apparent correlations, \mbox{$|r_{\rm S}|\approx 0.3$--0.6} indicates apparent correlations, and \mbox{$|r_{\rm S}|\approx 0.6$--1.0} indicates strong correlations.\label{spearman}} in Table~\ref{stat}).
This result does not appear to be a selection effect because it is also obtained when examining \mbox{Sample C} that is roughly complete for both blue and red galaxies.

\begin{figure*}[ht]
\includegraphics[width=7.1in]{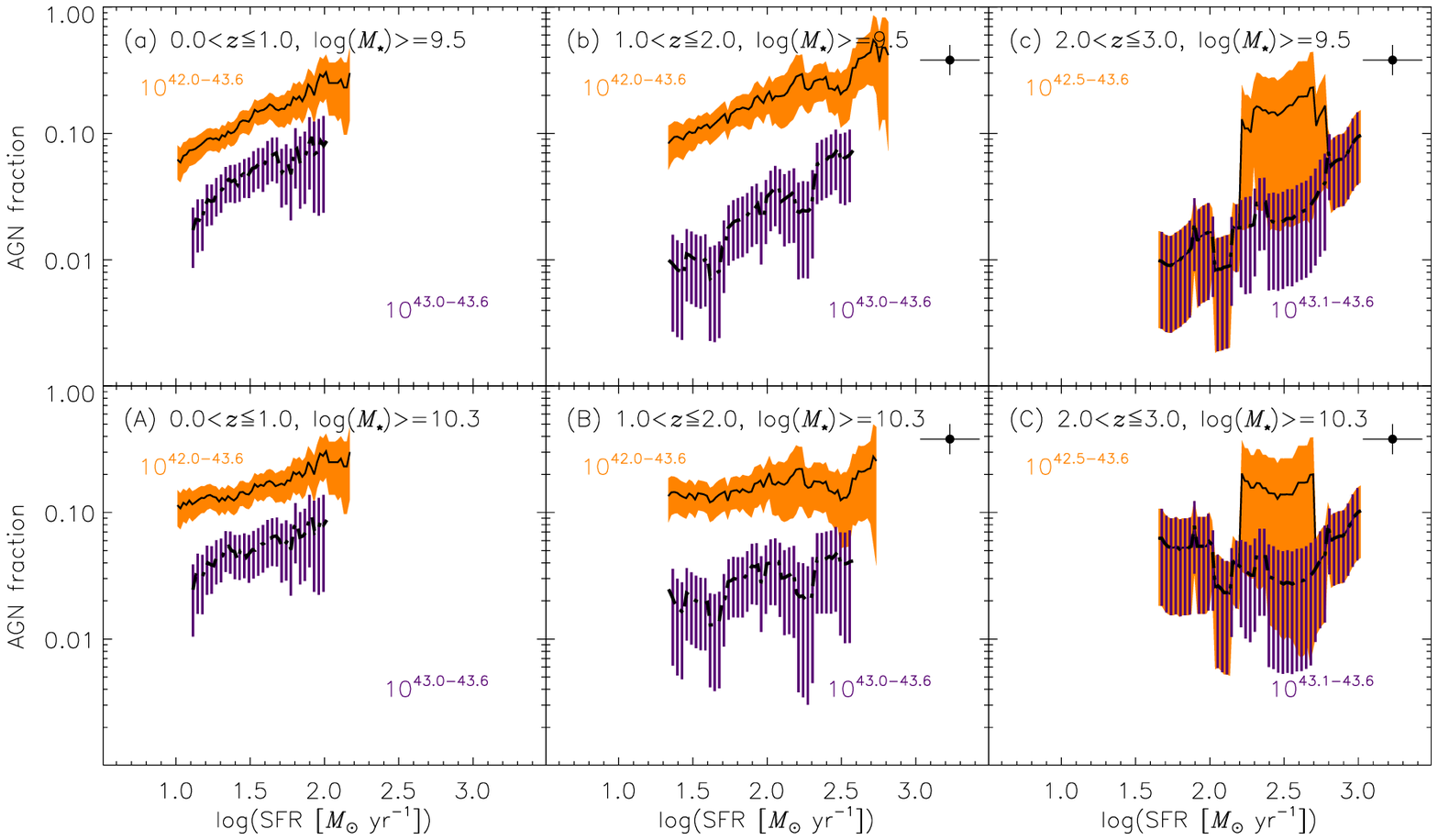}
\caption{AGN fraction as a function of SFR in bins of $\Delta {\rm log(SFR)}=0.5$ for \mbox{Sample B} (top panels) and \mbox{Sample C} (bottom panels), respectively (symbols have the same meaning as those in Fig.~\ref{cmd}).
A cut of ${\rm SFR}\geq 10$ ($\geq 20$, $\geq 30$) $M_\odot$ yr$^{-1}$ was applied for the redshift bin of $0.0<z<1.0$ ($1.0\leq z<2.0$, $2.0\leq z \leq 3.0$) in order to remove the uncertainties caused by upper limits on SFR (see \S~\ref{overview}). The filled circles indicate the approximate AGN fraction for $z\approx 2$ submillimeter galaxies (Alexander et al. 2005b). (A color version of this figure is available in the online journal.)}
\label{agn-sfr}
\end{figure*}

(3) From Fig.~\ref{cmd-z}(e), it seems that the AGN fraction rises toward higher SFRs (also see Rafferty et al. 2010). To examine this result further, we plot AGN fraction as a function of SFR in different ranges of redshift for \mbox{Sample B} and \mbox{Sample C} in Fig.~\ref{agn-sfr}.
For \mbox{Sample B}, we find that the fraction of AGNs generally rises as the SFR increases (note that at the high-SFR end this trend is roughly consistent with the AGN fraction for $z\approx 2$ SMGs obtained by Alexander et al. 2005b).
This trend persists if we consider the X-ray luminous AGNs [i.e., $43.0\leq {\rm log}(L_X/({\rm erg\:s^{-1}}))\leq 43.7$].
This result implies that host galaxies of \mbox{X-ray} AGNs generally have higher SFRs than non-AGN galaxies from \mbox{$z\approx 0$--3} for \mbox{Sample B}.
However, when examining \mbox{Sample C},
we find that the above trend of increasing AGN fraction toward higher SFRs 
becomes less prominent at $1<z\leq 2$ and is not detectable at $2<z\leq 3$, although this trend does exist at $0<z\leq 1$; 
similar results were obtained considering the X-ray luminous AGNs.
The decrease in the strength of the AGN fraction vs. SFR correlation with increasing redshift, at least from \mbox{$z\approx 0$--2}, appears to be a real effect not attributable to limited source statistics; this is discussed further in \S~\ref{mass-effect}. 

(4) The presence of about a dozen AGN hosts (with redshifts up to $\approx 1.3$ and a median value of $L_{\rm X}\approx 7\times 10^{42}$ \mbox{erg s$^{-1}$}) with very low SFRs (i.e., $\lsim 1.0 M_\odot$ yr$^{-1}$; see Fig.~\ref{cmd-z}(e)) implies that AGN activity may persist up to $z \approx 1.3$ after the process of active star formation is largely over, which agrees with previous works (e.g., Nandra et al. 2007; Georgakakis et al. 2008).

\begin{figure*}[ht]
\includegraphics[width=7.1in]{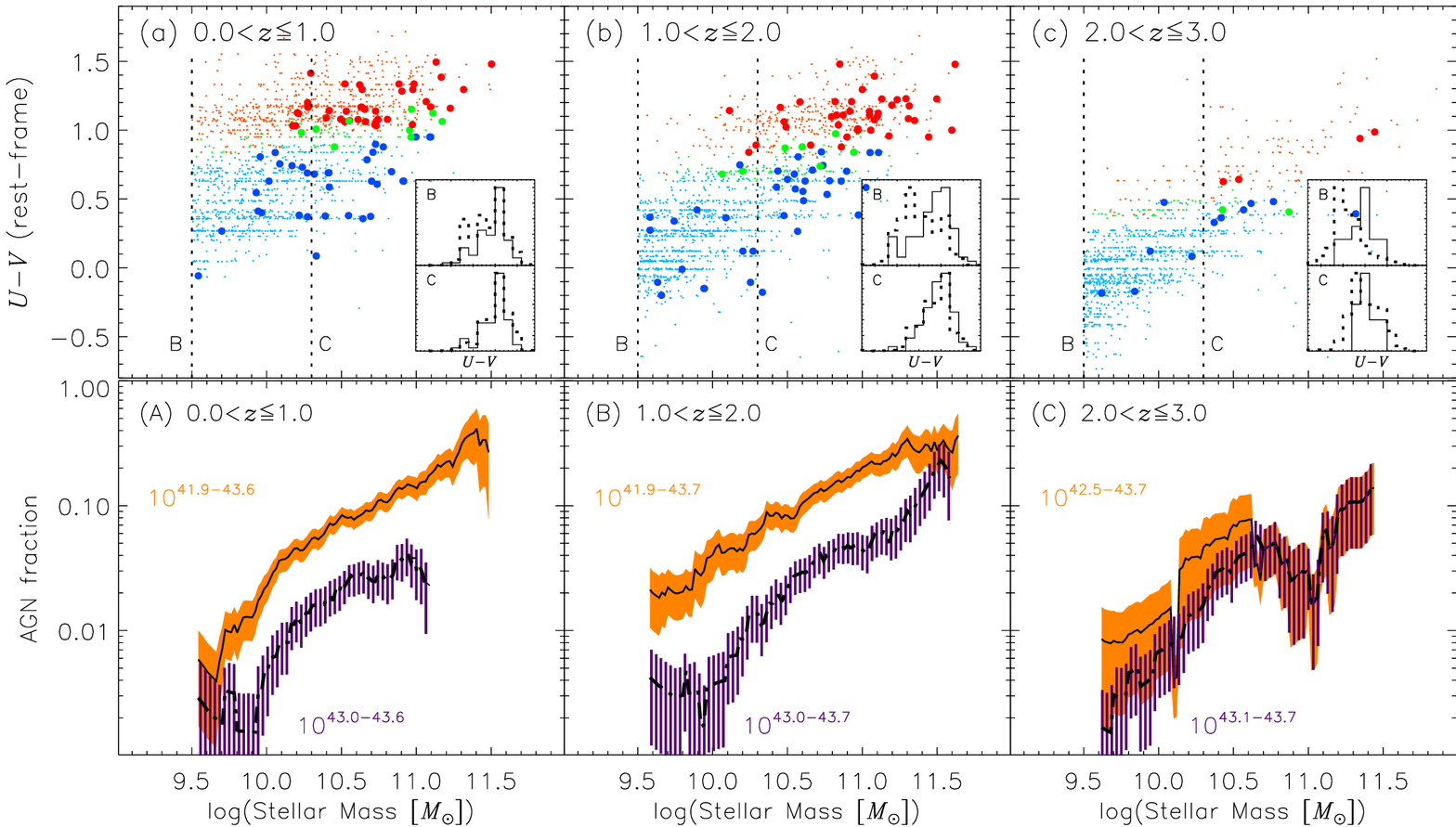}
\caption{(a -- c) Color-mass diagrams for \mbox{Sample B}:
AGN hosts are shown as filled circles and non-AGN galaxies are shown as small dots 
(colors have the same meaning as those in Fig.~\ref{cmd}(a -- i)).
The two sets of dotted lines indicate the stellar-mass cut for \mbox{Sample B} ($\geq 10^{9.5}M_\odot$) and \mbox{Sample C} ($\geq 10^{10.3}M_\odot$), respectively.
Shown in the insets are the histograms of rest-frame $U-V$ color for \mbox{Sample B} and \mbox{Sample C}, with AGN hosts shown as solid lines and non-AGN galaxies as dotted lines (the peak value of each histogram has been rescaled to unity).
(A -- C) AGN fraction as a function of stellar mass in bins of $\Delta {\rm log}(M_\star)=0.5$ for \mbox{Sample B}
(symbols have the same meaning as those in Fig.~\ref{cmd}(A -- I)). (A color version of this figure is available in the online journal.)}
\label{agn-red-mass}
\end{figure*}

(5) There is a strong color evolution as a function of stellar mass
such that the more massive the galaxy, the redder its color, for both AGN hosts and non-AGN galaxies 
(see Figs.~\ref{cmd-z}(h -- k) and the values of $r_{\rm S}$ in Table~\ref{stat}).
Furthermore, the AGN fraction seems to rise strongly toward higher stellar masses (see Fig.~\ref{cmd-z}(h)).
Figure~\ref{agn-red-mass} substantiates these two results, showing the rest-frame $U-V$ color and AGN fraction as a function of stellar mass in different ranges of redshift for \mbox{Sample B} (similar results were obtained considering \mbox{Sample C}).
First, both AGN hosts and non-AGN galaxies generally become redder as the stellar mass increases irregardless of redshift range considered.
This result agrees with previous works at lower redshifts (e.g., Baldry et al. 2006). 
The color histograms for both AGN hosts and non-AGN galaxies in \mbox{Sample B} and \mbox{Sample C} (shown in the insets) strengthen a result obtained earlier in this section:
for \mbox{Sample B}, AGN hosts generally appear redder than non-AGN galaxies; whereas for \mbox{Sample C}, AGN hosts and non-AGN galaxies have similar color distributions, no matter which range of redshift is considered.
Second, the fraction of AGNs increases strongly as the stellar mass increases irregardless of redshift range considered (also true for X-ray luminous AGNs), which agrees with the result obtained in \S~\ref{sec:cmd} that most AGNs reside in massive hosts
and extends previous findings to higher redshifts (e.g., Kauffmann et al. 2003; Best et al. 2005; Bundy et al. 2008; Alonso-Herrero et al. 2009; Brusa et al. 2009; Silverman et al. 2009).
We note that for a given lower limit on the AGN luminosity, 
we can identify lower Eddington ratio ($L/L_{\rm Edd}$) AGN activity in higher-mass galaxies than 
in lower-mass galaxies, which introduces an inherent bias  
toward the above trend of an increasing AGN fraction with increasing stellar mass. 
A careful investigation of this bias would be useful, but is beyond the scope of this work. 

\subsection{Mass-Selection Effects}\label{mass-effect}

In \S~\ref{sec:cmd}, \S~\ref{color-dep}, and Fig.~\ref{cmd}, we showed that results about color-magnitude relations of AGN hosts and non-AGN galaxies change when considering \mbox{Sample B} vs. \mbox{Sample C}, which have different stellar-mass cuts.
For example, the AGN fraction rises toward redder colors in \mbox{Sample B}, but the AGN fraction remains nearly constant regardless of color in \mbox{Sample C}.
We note that over the wide stellar-mass range that makes up \mbox{Sample B}
($M_\star\geq 10^{9.5}M_\odot$) there is significant stellar-mass dependent variation in the distribution of $U-V$ colors (see Fig.~\ref{cmd-z}(h)).
Since the AGN fraction increases
with stellar mass and the $U-V$ color becomes redder with increasing stellar
mass, we will artificially find that the AGN fraction is larger for redder galaxies.
Furthermore, the incompleteness for red galaxies at high redshifts in \mbox{Sample B} also contributes to this bias
because the inclusion of a large population of low-mass blue galaxies (due to a relatively low stellar-mass cut) makes the AGN fraction become smaller toward blue colors.  
In order to mitigate this bias, we have implemented a mass-matching technique. 
We constructed a mass-matched sample as follows:
for each AGN host in \mbox{Sample B}, ten unique galaxies (i.e., no duplicates) with a similar stellar mass (i.e., $M_{\star,{\rm AGN}}/2\leq M_{\star,{\rm GAL}}\leq 2M_{\star,{\rm AGN}}$) in \mbox{Sample B} were randomly selected (ignoring all information about their $M_{\rm V}$, $U-V$, and SFR values).
Note that \mbox{Sample C} happens to be a roughly mass-matched sample (see Fig.~\ref{cmd-z}(h)) that ensures a fair comparison between AGN hosts and non-AGN galaxies.

\begin{figure*}[ht]
\includegraphics[width=7.1in]{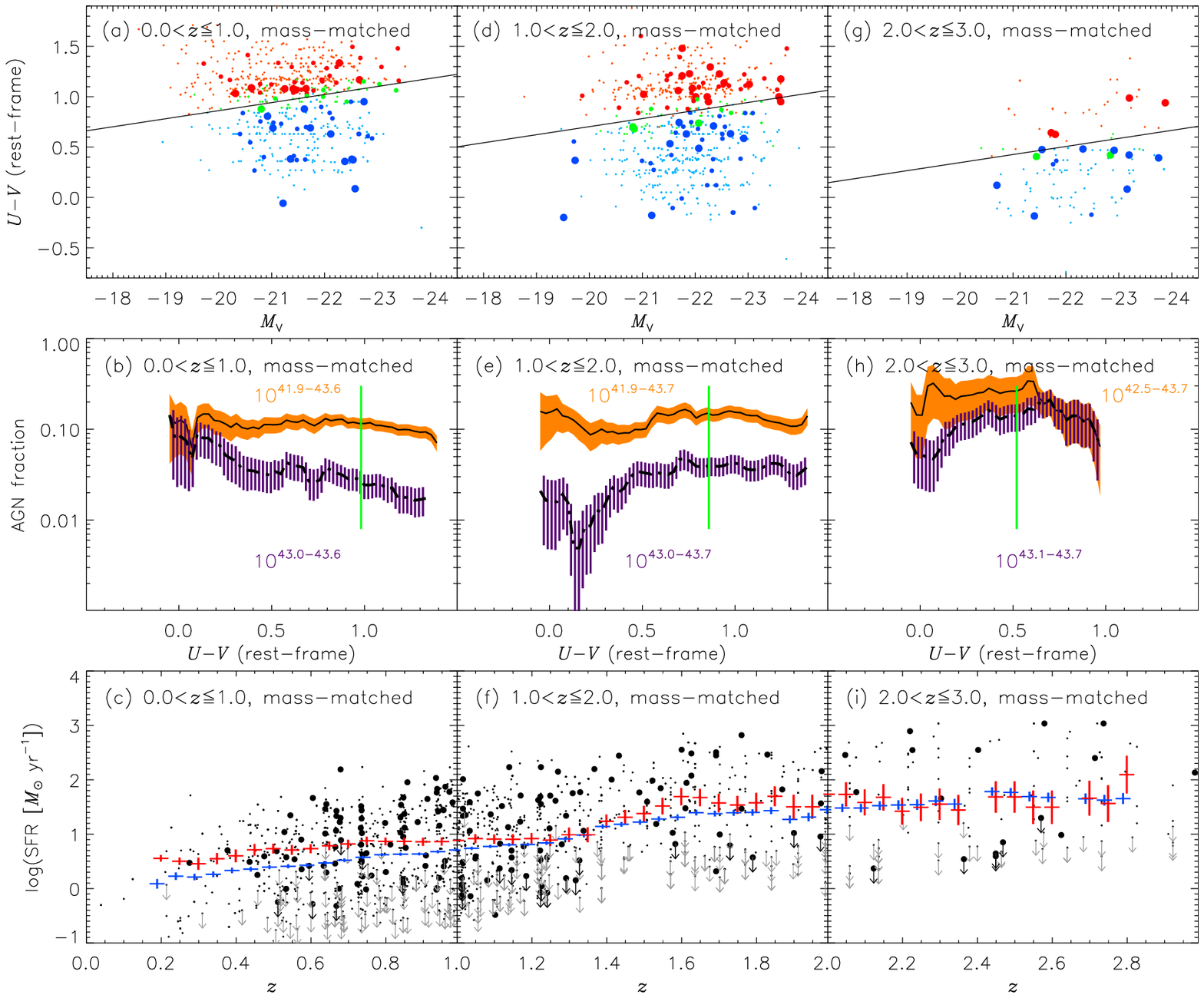}
\caption{(Top, Middle) Same as Fig.~\ref{cmd}, but for a mass-matched sample as defined in \S~\ref{mass-effect}.
(Bottom) SFR or upper limit on SFR as a function of redshift for AGN hosts (shown as filled circles) and non-AGN galaxies (small dots) in the above mass-matched sample (symbols have the same meaning as those in Fig.~\ref{cmd-z}(e)).
The red (blue) crosses show the running means and associated errors of SFR (in bins of $\Delta z=0.5$) for AGN hosts (non-AGN galaxies), which were computed using the Kaplan-Meier estimator provided by ASURV that treats censored data (see \S~\ref{mass-effect}).
The red crosses have been slightly shifted rightward for clarity of presentation.
(A color version of this figure is available in the online journal.)}
\label{fig:masseffect}
\end{figure*}

The top and middle panels of Fig.~\ref{fig:masseffect} show CMDs and AGN fractions as a function of color for such a mass-matched sample. 
Note that the absolute normalizations of the AGN fractions in Fig.~\ref{fig:masseffect} are meaningless due to the above random-drawing scheme of ten galaxies vs. one AGN, and we are only interested in trends with color here.
It seems clear, for a mass-matched sample, that 
there is no apparent clustering of AGNs in the CMDs and
the AGN fraction does not vary much with color up to $z\approx 2$--3.
We note that there appears to be a weak trend at $0.0<z\leq 1.0$ such that the fraction of \mbox{X-ray} luminous AGNs increases toward blue colors, which has also been seen in \mbox{Sample C} [see Fig.~\ref{cmd}(D)].
This weak trend, which is unlikely to be due to AGN blue light contamination as demonstrated in \S~\ref{agn-contamination}, implies a corresponding weak trend that 
the fraction of relatively low-luminosity AGNs (i.e., $41.9\leq {\rm log}(L_{\rm X}/({\rm erg\:s^{-1}}))\leq 43.0$) increases toward red colors.
These results confirm those obtained previously when considering \mbox{Sample C}.
We repeated the construction of such a random mass-matched sample many times and find that these conclusions are stable. 
As a further check,
we also constructed random non-mass-matched samples (i.e., for each AGN host, 10 unique galaxies with any stellar mass were randomly selected) 
and find that non-mass-matched samples do behave differently from mass-matched samples, i.e.: the results obtained with non-mass-matched samples
are different from those obtained with mass-matched samples, but are similar to those obtained with \mbox{Sample B} that is essentially non-mass-matched. 
This clearly demonstrates the mass-selection biases associated with non-mass-matched samples.

\begin{figure*}[ht]
\includegraphics[width=7.1in]{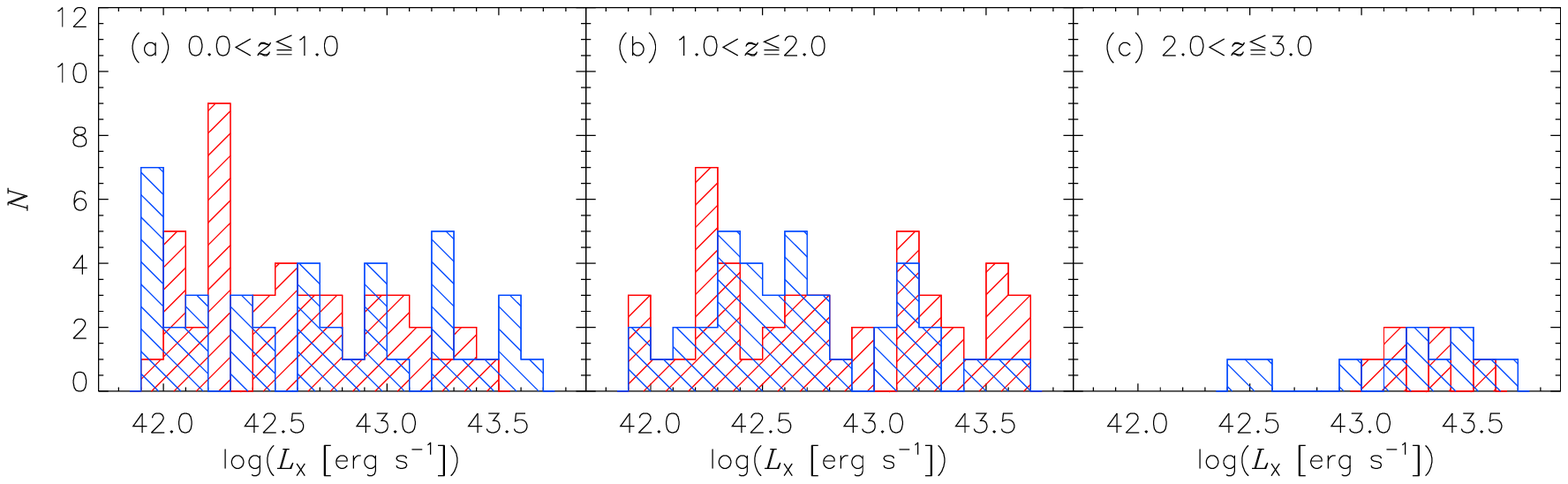}
\caption{Histograms of X-ray luminosity for AGN hosts in Sample B.
The red (blue) histograms are for red (blue) AGN hosts (red and blue AGN hosts are separated using Eqn.~(\ref{separation-line}) in \S~\ref{sec:bi-gal}). The two histograms in each panel are statistically indistinguishable.
(A color version of this figure is available in the online journal.)}
\label{lx-histo}
\end{figure*}

Since the AGN fraction is approximately constant across colors for mass-matched samples,
it is of interest to see whether the level of AGN activity is similar across colors.
To this end, we plot in Fig.~\ref{lx-histo} histograms of \mbox{X-ray} luminosity (a proxy for the level of SMBH accretion) 
for AGNs with blue and red hosts in \mbox{Sample B}.
We find that AGNs with blue and red hosts have similar distributions of \mbox{X-ray} luminosity (confirmed by K-S tests),
which indicates that AGNs with blue and red hosts have a comparable level of SMBH accretion (a similar result was obtained considering \mbox{Sample C}). 

The bottom panels of Fig.~\ref{fig:masseffect} show 
plots of SFR or upper limit on SFR as a function of redshift for AGN hosts and non-AGN galaxies in the above mass-matched sample.
To make sample comparisons with upper limits included, we utilized the Astronomy SURVival Analysis package (ASURV; Isobe et al. 1986; Lavalley et al. 1992)
to compute sample means using the Kaplan-Meier (K-M) estimator
and four nonparametric sample-comparison tests for censored data: the Gehan's generalized Wilcoxon test,
the logrank test, the Peto \& Peto generalized Wilcoxon test,
and the Peto \& Prentice generalized Wilcoxon test
(these four tests typically produce similar results).
We find, according to sample means derived using the K-M estimator as well as results from the above nonparametric sample-comparison tests, that 
(1) the SFRs of both AGN hosts and non-AGN galaxies generally increase toward higher redshifts from \mbox{$z\approx 0$--3}, which is consistent with the well-established formation epoch at \mbox{$z\approx 1$--4}; and
(2) at $0<z\leq 1$ the SFRs of AGN hosts are generally a factor of \mbox{$\approx 2$--3} [i.e., \mbox{$\approx 0.3$--0.4} dex; see Fig.~\ref{fig:masseffect}(c)] larger than those of non-AGN galaxies, while at $1<z\leq 2$ and $2<z\leq 3$ AGN hosts and non-AGN galaxies on average have similar SFRs (i.e., the difference between their means of SFR is typically within \mbox{0.1--0.2} dex).
The latter results are in agreement with previous results on AGN fraction vs. SFR obtained with \mbox{Sample C} in \S~\ref{color-dep}.
Our result that AGN hosts have elevated SFRs at $0<z\leq 1$ also confirms the result from Silverman et al. (2009)
who obtained a similar ratio between the SFRs of AGN hosts and non-AGN galaxies at $0<z\leq 1$ [see their Fig.~9(f)].
%

We also find, if considering star-forming populations only [i.e., applying
a cut of ${\rm SFR}\geq 10$ ($\geq 20$, $\geq 30$) $M_\odot$ yr$^{-1}$ to both AGN hosts and non-AGN galaxies
at $0.0<z<1.0$ ($1.0\leq z<2.0$, $2.0\leq z \leq 3.0$)], 
that both AGN hosts and non-AGN galaxies have similar SFR distributions irregardless of redshift range considered. 
This result may contribute to explaining the 
interesting trend we mentioned above that the factor of \mbox{$\approx 2$--3} difference between the SFRs of AGN hosts and non-AGN galaxies diminishes above $z\approx 1$.
The majority of AGN hosts/non-AGN galaxies at $z\gsim 1$ have a SFR greater than \mbox{$\approx 5$--10 $M_\odot$ yr$^{-1}$},
which means that their SFR distributions are dominated by star-forming populations (thus being similar).
Further investigation is required to understand the physical origin of 
AGN hosts having higher SFRs than non-AGN galaxies at low redshifts vs. AGN hosts and non-AGN galaxies having similar SFRs at high redshifts.

There might appear to be an inconsistency between the two results obtained with mass-matched samples at $0<z\leq 1$ that AGN hosts have higher SFRs than non-AGN galaxies and that the color distributions of AGN hosts and non-AGN galaxies are similar, since the former result suggests that AGN hosts should have bluer colors than non-AGN galaxies.
However, as shown in \S~\ref{color-dep}, the correlation between colors and SFRs is not very tight:
for a given color, the SFRs span a wide range, although there may be a rough trend of bluer colors toward higher SFRs. 
Furthermore, the SFRs of AGN hosts are roughly a factor of \mbox{2--3} larger than the SFRs of non-AGN galaxies at $0<z\leq 1$, which does not make a strong difference in colors.
Therefore, there is no real inconsistency in this issue. 

\section{Conclusions and Summary}\label{summary}

In this work, we have assembled a sample of X-ray-selected moderate-luminosity AGNs as well as their parent galaxies
in the most sensitive central areas of the \mbox{2~Ms} CDFs 
to extend color-magnitude relations of AGN hosts and non-AGN galaxies through the galaxy formation epoch.
The uniqueness of this sample allows us to explore several important issues up to \mbox{$z\approx 2$--3} when most of galaxy assembly occurred;
such explorations at high redshifts have been largely unattainable in previous studies primarily due to data-depth issues.
The main advantages of this work are
(1) we have used the deepest \mbox{X-ray} data, the best multiwavelength coverage, and spectroscopic/photometric redshifts of the highest quality to date; and
(2) we have used a template SED fitting method to derive galaxy physical properties (e.g., colors, stellar masses, and SFRs), which significantly reduces the failure rate of estimations for high-redshift sources whose SEDs are more prone to large uncertainties or errors.
We summarize the main results of this work as follows: 

\begin{enumerate}

\item Non-AGN galaxy color bimodality (i.e., existence of the red sequence and the blue cloud) exists up to \mbox{$z\approx 3$}, with or without correcting for dust extinction.
However, there appears to be no apparent color bimodality for AGN hosts up to \mbox{$z\approx 2$}.
These results (see \S~\ref{sec:bi}) confirm previous works and extend them to higher redshifts.    

\item For mass-matched samples up to \mbox{$z\approx 2$--3} (e.g., \mbox{Sample C}):
(1) both AGN hosts and non-AGN galaxies are luminous and share similar $M_{\rm V}$ distributions;
(2) AGN clustering in the CMD is not apparent; and
(3) the AGN fraction remains nearly constant (at \mbox{$\approx 10$\%}) regardless of color,
which likely indicates the duty cycle of SMBH growth in typical massive galaxies
(see \S~\ref{sec:cmd} -- \S~\ref{mass-effect}).
These results are in contrast to those obtained with 
non-mass-matched samples up to \mbox{$z\approx 2$--3} (e.g., \mbox{Sample B/D}):
(1) AGN hosts generally appear more luminous (i.e., having more negative values of $M_{\rm V}$) than non-AGN galaxies;
(2) AGNs predominantly reside in the red sequence, the top of the blue cloud, and the green valley in between (i.e., AGN clustering in the CMD); and
(3) the AGN fraction generally increases as the color becomes redder (see \S~\ref{sec:cmd} and \S~\ref{color-dep}).  

\item Luminous SMGs, the prototypical examples of extreme star-forming galaxies at \mbox{$z\approx 1$--3},
are spread throughout the CMD (see \S~\ref{sec:cmd}). 
Likely due to dust-extinction effects, they often do not show the blue colors typically associated with star-forming galaxies. This underscores the need for caution when studying the color-magnitude properties of small samples of remarkable sources.

\item Most AGNs reside in massive hosts (i.e., $M_\star\geq 10^{10.3}M_\odot$),
and the AGN fraction rises strongly toward higher stellar masses (see \S~\ref{color-dep}), up to \mbox{$z\approx 2$--3},
which confirms and extends previous works at lower redshifts. 

\item There are some strong trends of non-AGN galaxy/AGN host rest-frame colors, up to \mbox{$z\approx 2$--3} (see \S~\ref{color-dep}): 
(1) the colors of both non-AGN galaxies and AGN hosts generally become bluer as the redshift increases; and
(2) the colors of both non-AGN galaxies and AGN hosts generally become redder as the stellar mass increases.

\item For mass-matched samples, the SFRs of AGN hosts are on average a factor of \mbox{$\approx 2$--3} larger than those of non-AGN galaxies at \mbox{$z\approx 0$--1}, but this difference diminishes at \mbox{$z\approx 1$--3} (see \S~\ref{mass-effect}).

\end{enumerate}

It has been demonstrated that galaxy evolution is best probed using stellar-mass selected samples
(e.g., van Dokkum et al. 2006; Kriek et al. 2008; and references therein).
van Dokkum et al. (2006) recommend that samples of high-redshift galaxies be selected by stellar mass, rather than by color or luminosity, 
because (1) the mass evolution of galaxies is probably gradual, whereas luminosities and colors can vary dramatically on short timescales due to
starbursts and dust, and (2) models of galaxy formation can predict masses with higher confidence than luminosities and colors.
Moreover, stellar-mass selected samples make comparisons between the properties of AGN hosts and non-AGN galaxies more appropriate.
For example, Silverman et al. (2009) showed that
a mass-selected sample is required to reduce the strength of an artificial peak in the AGN fraction falling in the transition
region (i.e., the green valley) due to the fact that many blue-cloud galaxies have low mass-to-light ratios in luminosity limited samples.

In this work, we have further demonstrated the importance of using mass-matched samples  
for investigations of color-magnitude relations of active and non-active galaxies, 
because results and conclusions obtained with a mass-matched sample are different from those obtained with a non-mass-matched sample 
(see \S~\ref{sec:cmd} -- \S~\ref{mass-effect}).
For instance, for a non-mass-matched sample,
AGNs predominantly reside in the red sequence, the top of the blue cloud, and the green valley in between;
this has been presented as evidence for AGN feedback because the location of AGN hosts in the CMD
is consistent with the transitional region where the blue-to-red migration of galaxies occurs due to the quenching of star formation (e.g., Nandra et al. 2007).
However, the AGN clustering in the CMD largely disappears and the AGN fraction remains nearly constant 
as a function of color when considering a mass-matched sample. 
Therefore, in order to obtain a complete picture about the role of AGNs in galaxy evolution by means of CMD analyses, 
one must fully take into account the dependence of AGN activity on stellar mass (i.e., using mass-matched samples). 

With mass-selection effects taken into account, we find that the main results in this work 
(except for color-bimodality, which likely needs some quenching mechanism)
can be reasonably explained by a combination of two main ingredients: (1) AGNs at \mbox{$z\approx 0$--3} preferentially reside in massive galaxies that
generally tend to have redder colors (i.e., color-mass correlation);
and (2) galaxies evolve passively (e.g., Kodama \& Arimoto 1997; Stanford, Eisenhardt, \& Dickinson 1998; van Dokkum \& Franx 2001; Daddi et al. 2004; Bothwell, Kennicutt, \& Lee 2009) or secularly (e.g., Kormendy \& Kennicutt 2004; Hopkins \& Hernquist 2006); this ingredient is mainly responsible for the evolution of galaxies, i.e., blue-to-red migration.
Consequently, our results tightly constrain any effects from moderate-luminosity AGN feedback upon the color-magnitude properties of galaxies from \mbox{$z\approx 0$--3} when most of galaxy assembly occurred. 
We note that moderate-luminosity AGNs, like those studied here, dominate 
the accretion density and SMBH growth only at \mbox{$z\approx 0$--1} 
while more luminous AGNs dominate above $z\approx 1$ (e.g., Ueda et al. 2003; Hasinger, Miyaji, \& Schmidt 2005). 
For moderate-luminosity AGNs, some simulations predict that feedback may not be important (e.g., Hopkins \& Hernquist 2006), whereas other models suggest it is (e.g., Fabian, Vasudevan, \& Gandhi 2008; Raimundo et al. 2010).
Given that such simulations and models generally rely on significant assumptions and are therefore subject to large uncertainties, further observational tests of 
moderate-luminosity AGN feedback are required.

We show that the AGN fraction is approximately constant across colors for galaxies with similar stellar masses
and that AGNs with blue and red hosts have comparable \mbox{X-ray} luminosities (see \S~\ref{mass-effect}). 
These results imply that AGN activity is about as prevalent in massive blue galaxies as it is in red galaxies.
Assuming that the SMBH mass generally decreases toward blue colors since the bulges may be
smaller in blue (late-type) hosts than in red (early-type) hosts for a given galaxy stellar mass,
the less-massive SMBHs in blue hosts may undergo more significant fractional growth
(i.e., with larger Eddington ratios) than those in red hosts.
If this effect exists and is significant, 
AGN feedback may be more likely to occur in those systems having higher ratios of $L_{\rm AGN}/M_{\rm bulge}$.
However, the above assumption is not certain.
For example, Schawinski et al. (2010) show that in local universe
there is no simple relationship between colors and morphological types
[e.g., a red (blue) galaxy is not necessarily an early-type (late-type) galaxy and vice versa]; they also show that
both early- and late-type AGN hosts have similar typical SMBH 
masses and Eddington ratios (although both SMBH masses and Eddington ratios span a broad range).
At high redshifts, it is more challenging to obtain reliable morphological information and estimates of SMBH masses. 
Therefore, it is difficult to assess how significant AGN feedback effects could be in some massive blue galaxies 
if such feedback even exists at \mbox{$z\approx 0$--3},
given the current constraints.
Further investigation of these effects would be worthwhile.

A great deal of further work, in addition to that already mentioned,
can be done to improve  and extend the analyses presented here, e.g., 
obtaining even deeper \mbox{X-ray} and near-infrared data.
According to Fig.~\ref{lum-z}(a), we would expect to identify an AGN with $L_{\rm X}\gsim 10^{43}$ \mbox{erg s$^{-1}$} out to $z\approx 3$,
but we would not expect to identify an AGN with $L_{\rm X}$ a few times $10^{42}$ \mbox{erg s$^{-1}$} beyond $z\approx 1.5$.
We note that this is already the best that the currently deepest \mbox{X-ray} surveys, the \mbox{2~Ms} CDFs, can offer.
Therefore, \mbox{X-ray} data with even higher sensitivity are required to provide better completeness, 
which will also increase the size of the AGN sample by detecting more highly obscured and moderate-luminosity AGNs at high redshifts.
%
%
Obtaining ultradeep near-infrared data would also be beneficial for the analyses here.
First, ultradeep near-infrared selected source catalogs would include a population of less-massive red galaxies and thus lower the stellar-mass completeness limit.
Second, ultradeep near-infrared data have proved critical to improve $z_{\rm zphot}$ estimates and thus estimates of other source properties, especially for high-$z$ sources, e.g., Wardlow et al. (2010) made use of ultradeep \mbox{HAWK-I} $J$ and $K_{\rm s}$ data as well as other multiwavelength data to derive high-quality photometric redshifts for a sample of \mbox{E-CDF-S} SMGs. 
Another valuable project within the framework of color-magnitude and color-mass relations, as an extension of our work here,
would be to constrain the colors of quasar hosts out to high redshifts, although practically this is difficult because the host light is overwhelmed by quasar emission except for highly obscured quasars.

\acknowledgments

Support for this work was provided by NASA through {\it Chandra} Awards SP8-9003A 
(YQX, WNB, BL, and DAR) and SP8-9003B (FEB) issued by the {\it Chandra} \mbox{X-ray} Observatory Center, which 
is operated by the Smithsonian Astrophysical Observatory. 
We acknowledge financial support from NASA ADP grant NNX10AC99G (YQX, WNB),
the Royal Society (DMA), the Leverhulme Trust (DMA),
the Science and Technology Facilities Council fellowship program (BDL), and
the Einstein Postdoctoral Fellowship grant PF9-00064 (BDL).
We thank the referee for constructive feedback.
We also thank R. Feldmann for helpful discussions on the use of the ZEBRA code
and J. Wu for help with using the ASURV package.

\end{document}